\documentclass[acmsmall,screen]{acmart}

\newcommand{\sysname}{{\textsc{MICRYSCOPE}}\xspace}

\usepackage[most]{tcolorbox}
\usepackage{tikz}
\usepackage{amsmath}
\usepackage{amssymb} 

\usepackage{subcaption}
\usepackage{wrapfig}
\usepackage[labelfont=bf]{caption}
\usepackage{booktabs}
\usepackage{makecell}

\usepackage{pifont}
\usepackage{colortbl}
\usepackage{xcolor}
\usepackage{tabularx}
\usepackage[draft]{minted} 

\usepackage{svg}

\usepackage{textcomp}
\usepackage{listings}
\usepackage{multirow}
\usepackage{appendix}

\usepackage{color}
\usepackage{soul}

\usepackage{algpseudocode}
\usepackage[linesnumbered,ruled,vlined]{algorithm2e}

\usepackage{pifont}
\newcommand{\cmark}{\checkmark}
\newcommand{\xmark}{\ding{55}}

\definecolor{backgroundcolor}{RGB}{245, 245, 245}
\definecolor{framecolor}{RGB}{200, 200, 200}
\setminted{
	bgcolor=backgroundcolor,
	frame=none,
	framesep=3mm,
	rulecolor=\color{framecolor},
	breaklines=true,
	breakanywhere=true,
	fontsize=\footnotesize
}

\microtypecontext{spacing=nonfrench}

\renewcommand{\paragraph}[1]{\noindent\textbf{#1}}

\providecommand{\warn}[1]{}  
\providecommand{\nop}[1]{}

\RequirePackage{etoolbox}

\makeatletter
\patchcmd{\hyper@makecurrent}{%
	\ifx\Hy@param\Hy@chapterstring
	\let\Hy@param\Hy@chapapp
	\fi
}{%
	\iftoggle{inappendix}{
		\@checkappendixparam{chapter}%
		\@checkappendixparam{section}%
		\@checkappendixparam{subsection}%
		\@checkappendixparam{subsubsection}%
		\@checkappendixparam{paragraph}%
		\@checkappendixparam{subparagraph}%
	}{}%
}{}{\errmessage{failed to patch hyper@makecurrent}}

\newcommand*{\@checkappendixparam}[1]{%
	\def\@checkappendixparamtmp{#1}%
	\ifx\Hy@param\@checkappendixparamtmp
	\let\Hy@param\Hy@appendixstring
	\fi
}
\makeatother

\newtoggle{inappendix}
\togglefalse{inappendix}

\apptocmd{\appendix}{\toggletrue{inappendix}}{}{\errmessage{failed to patch appendix}}
\apptocmd{\subappendices}{\toggletrue{inappendix}}{}{\errmessage{failed to patch subappendices}}

\newtcolorbox[auto counter]{mybox}[2][]{%
	enhanced jigsaw,
	colback=white!12,
	breakable,
	#1}

\date{}

\title{``MCP Does Not Stand for Misuse Cryptography Protocol'': Uncovering Cryptographic Misuse in Model Context Protocol at Scale}

\author{Biwei Yan}
\affiliation{
	\institution{Shandong University}
	\country{China}
}
\email{bwyan@sdu.edu.cn}

\author{Yue Zhang}
\affiliation{
	\institution{Shandong University}
	\country{China}
}
\email{zyueinfosec@sdu.edu.cn}

\author{Minghui Xu}
\affiliation{
	\institution{Shandong University}
	\country{China}
}
\email{mhxu@sdu.edu.cn}

\author{Hao Wu}
\affiliation{
	\institution{Nanjing University}
	\country{China}
}
\email{hao.wu@nju.edu.cn}

\author{Yechao Zhang}
\affiliation{
	\institution{Shandong University}
	\country{China}
}
\email{yech.zhang@gmail.com}

\author{Kun Li}
\affiliation{
	\institution{Shandong University}
	\country{China}
}
\email{kunli@sdu.edu.cn}

\author{Guoming Zhang}
\affiliation{
	\institution{Shandong University}
	\country{China}
}
\email{guomingzhang@sdu.edu.cn}

\author{Xiuzhen Cheng}
\affiliation{
	\institution{Shandong University}
	\country{China}
}
\email{xzcheng@sdu.edu.cn}

\begin{document}
	
	\begin{abstract}
		The Model Context Protocol (MCP) is rapidly emerging as the middleware for LLM-based applications, offering a standardized interface for tool integration. However, its built-in security mechanisms are minimal: while schemas and declarations prevent malformed requests, MCP provides no guarantees of authenticity or confidentiality, forcing developers to implement cryptography themselves. Such ad hoc practices are historically prone to misuse, and within MCP they threaten sensitive data and services. We present \sysname, the first domain-specific framework for detecting cryptographic misuses in MCP implementations. \sysname combines three key innovations: a cross-language intermediate representation that normalizes cryptographic APIs across diverse ecosystems, a hybrid dependency analysis that uncovers explicit and implicit function relationships (including insecure runtime compositions orchestrated by LLMs) and a taint-based misuse detector that tracks sensitive data flows and flags violations of established cryptographic rules. Applying \sysname to 9,403 MCP servers, we identified 720 with cryptographic logic, of which 19.7\% exhibited misuses. These flaws are concentrated in certain markets (e.g., Smithery Registry with 42\% insecure servers), languages (Python at 34\% misuse rate), and categories (Developer Tools and Data Science \& ML accounting for over 50\% of all misuses). Case studies reveal real-world consequences, including leaked API keys, insecure DES/ECB tools, and MD5-based authentication bypasses.  Our study establishes the first ecosystem-wide view of cryptographic misuse in MCP and provides both tools and insights to strengthen the security foundations of this rapidly growing protocol.
	\end{abstract}
	
	\maketitle
	
	\section{Introduction}
 
The rapid adoption of Model Context Protocol (MCP)\cite{wiki_mcp} has reshaped how large language models (LLMs) interact with external tools, databases, and services. By decoupling model reasoning from tool execution, MCP provides a standardized interface for heterogeneous capabilities, enabling structured requests that are mediated by clients and executed by servers. This abstraction simplifies integration and creates a uniform communication channel, positioning MCP as a foundational layer in AI ecosystems. Recent developments further underscore its momentum:  Microsoft has also positioned MCP as the ``USB-C for AI apps,'' embedding it into its Windows AI Foundry to enable cross-platform interoperability~\cite{verge_microsoft}. Industry analyses suggest that MCP is on track to become the de facto middleware for intelligent systems, akin to HTTP in the web era, with forecasts indicating that more than 75\% of enterprises will invest MCP by 2025~\cite{superagi_trends}. These trends highlight not only MCP's accelerating adoption~\cite{axios_mcp,contentful_mcp} but also its growing role as critical infrastructure in the agentic AI ecosystem. \looseness=-1

Despite its promise, MCP’s built-in security mechanisms remain minimal. At the protocol level, it enforces JSON-based schemas, capability declarations, and identifier traceability. While these safeguards help prevent malformed requests and enable basic auditing, they fall short in two critical areas: authenticity and confidentiality. MCP cannot natively guarantee that request–response messages are genuine or protected from interception, forcing developers to implement custom cryptographic safeguards such as encryption, authentication codes, or digital signatures.

Here lies the crux of the problem. History shows that when security-critical tasks are delegated to individual developers, cryptographic misuse is not the exception but the norm~\cite{ami2022crypto,fischer2024challenges,karimova2025model,mandal2024belt, kim2024privacy}. From hard-coded keys~\cite{ChenCWWJSXP25,ShiYZ0Y0025} and fixed random seeds to weak hashes like MD5~\cite{torres2023runtime,wickert2022fix} and insecure modes such as ECB~\cite{chen2024towards}, decades of research have consistently demonstrated the prevalence of errors when developers directly handle cryptographic APIs. Within MCP, the risk is magnified: servers often mediate access to sensitive data and proprietary services, meaning that even subtle misuses such as predictable random values or unauthenticated encryption can escalate into systemic vulnerabilities with far-reaching consequences. \looseness=-1

Unfortunately, existing program analysis and misuse detection tools are ill-suited to the unique challenges of MCP. First, MCP servers are highly heterogeneous, spanning over ten programming languages (e.g., Python, C++, Java, JavaScript) with differing type systems, memory models, and cryptographic libraries.  Second, MCP tools are often weakly coupled, leaving it to the LLM at runtime to orchestrate function compositions that may combine otherwise benign operations into insecure workflows. For instance, a developer may provide one function that derives a key by simply truncating a password and another that encrypts data using AES-CBC. Individually, both functions appear reasonable (but when the LLM chains them together in response to a user prompt like ``\textit{encrypt this file with my password,}''  the result is a dangerously weak encryption scheme). Third, even when the cryptographic APIs themselves are recognized, the real challenge lies in understanding intent. The same MD5 function may be perfectly acceptable when checking for accidental file corruption, yet dangerously inadequate when used for password storage. In other words, detecting misuse in MCP requires reasoning about semantics and context, not just pattern matching. \looseness=-1

To address these challenges, we present \sysname, a domain-specific analysis framework for detecting cryptographic misuses in MCP implementations. \sysname introduces three key innovations: (i) a cross-language intermediate representation (IR) that normalizes cryptographic API usage across diverse ecosystems, ensuring consistent detection across languages; (ii) a hybrid dependency analysis that reconstructs both explicit and implicit relationships among functions in MCP’s plugin-style architecture. Unlike traditional control-flow analysis, this approach models \textit{may-dependencies} created at runtime when the LLM orchestrates weakly coupled functions, thereby exposing insecure compositions that static analysis alone cannot reveal; and (iii) a taint-based misuse detector that tracks data flows across function boundaries and flags violations of well-established cryptographic security rules.

Through a large-scale study of 9,403 MCP servers, \sysname identified 720 servers that implemented cryptographic logic, of which 19.7\% exhibited misuses. These vulnerabilities are not evenly distributed: the Market Smithery Registry shows the highest density of insecure servers (42\%), while Mcpmarket, though the largest platform, contains a lower proportion of misuses (37\%). At the language level, Python servers (34\%) are disproportionately prone to errors compared to other languages, reflecting the uneven quality of its cryptographic library ecosystem. From a functional perspective, Developer Tools and Data Science \& ML categories account for more than 50\% of all misuses, underscoring that critical flaws emerge in the very tools developers depend on most.  Beyond statistics, our case studies highlight tangible consequences: hard-coded LLM API keys that attackers could immediately exploit for financial abuse, DES in ECB mode exposed through MCP tools that propagate insecure primitives into downstream applications, and MD5-based authentication tokens that allow adversaries to bypass deployment pipelines.  ``MCP''  now being read as ``\textit{Misuse Cryptography Protocol}'' rather than ``\textit{Model Context Protocol}''.

In summary, this work makes the following contributions:
\begin{itemize}
    \item  We provide the first systematic study of cryptographic misuse in MCP, highlighting its causes, patterns, and consequences.
    \item We design \sysname, a scalable analysis framework that integrates a cross-language intermediate representation, hybrid dependency reasoning, and taint-based detection to identify MCP-specific cryptographic misuses that existing tools fail to capture.
    \item We analyze 9,403 MCP servers and find that 19.7\% of crypto-enabled servers contain misuses, mapping their prevalence across markets, languages, and categories, and demonstrating real-world consequences through case studies that expose financial, confidentiality, and integrity risks.
\end{itemize}

	\begin{figure*}[t]
    \centering
    \tiny
    \setlength{\fboxsep}{0pt}  

    \newcommand{\codefontsize}{\fontsize{6}{7}\selectfont}

    \begin{minipage}[t]{0.31\textwidth}
        \centering
        \caption{\scriptsize LLM generates requests.}
        \label{lst:llm}
        \begin{minted}[fontsize=\codefontsize, breaklines, linenos, numbersep=3pt]{python}
user_query = "What are the top 3 students by credits?"

if "credits" in user_query.lower():
    tool_spec = {
        "tool": "sql.query",
        "params": {
            "sql": "SELECT name, credits FROM students "
                   "ORDER BY credits" 
                   "DESC LIMIT 3"
        }
    }
else:
    tool_spec = None
        \end{minted}
    \end{minipage}
    \hfill
    \begin{minipage}[t]{0.31\textwidth}
        \centering
        \caption{\scriptsize Clients translate requests.}
        \label{lst:client}
        \begin{minted}[fontsize=\codefontsize, breaklines, linenos, numbersep=3pt]{python}
import uuid, requests
SERVER = "http://127.0.0.1:8000/mcp"
msg = {
    "id": str(uuid.uuid4()),
    "version": "0.1",
    "type": "invoke",
    "tool": "sql.query",
    "params": {"sql": "SELECT name, credits "
                      "FROM students ORDER BY credits DESC LIMIT 3"}
}
resp = requests.post(SERVER, json=msg).json()
        \end{minted}
    \end{minipage}
    \hfill
    \begin{minipage}[t]{0.31\textwidth}
        \centering
        \caption{\scriptsize MCP Servers execute requests.}
        \label{lst:server}
        \begin{minted}[fontsize=\codefontsize, breaklines, linenos, numbersep=3pt]{python}
from fastapi import FastAPI, Request
import sqlite3

app = FastAPI()

@app.post("/mcp")
async def mcp_endpoint(req: Request):
    data = await req.json()
    sql  = data["params"]["sql"]
    con  = sqlite3.connect("students.db")
    cur  = con.cursor()
    cur.execute(sql)
    rows = cur.fetchall()
    return {"id": data["id"], "ok": True,
            "result": {"rows": rows}}
        \end{minted}
    \end{minipage}

\end{figure*}


\section{Background}

\subsection{Model Context Protocol Workflow}
 
Model Context Protocol (MCP) has been introduced as a standardized interface for enabling large language models to interact with external tools, data sources, and services in a secure and uniform manner. By decoupling model reasoning from tool execution, MCP establishes a context-sharing mechanism that allows models to invoke heterogeneous capabilities without relying on ad-hoc integration. The key components and their interactions can be summarized as follows:\looseness=-1

\begin{itemize}
  
    \item \textbf{LLM}: The large language model receives the user’s input (e.g., the user may ask ``\textit{What are the top 3 students by credits?}'') and first attempts to generate a response internally. When the task requires knowledge or capabilities beyond its training data (such as retrieving up-to-date information or executing a database lookup), the LLM produces a structured request and hands it off to the MCP client. \autoref{lst:llm} shows how the LLM decides to issue a standardized request to the \texttt{sql.query} tool (In MCP, a tool is an externally exposed capability  such as database query identified by a unique name and a defined input/output schema that the LLM can invoke through the client), instead of producing an answer solely from its internal knowledge.

    \item \textbf{MCP Client}: The client serves as the mediator between the LLM and the broader MCP ecosystem. It translates the LLM’s structured request into the MCP protocol format, attaches metadata such as request identifiers and versioning, and sends it to the designated MCP server. This process ensures that the LLM never directly interacts with heterogeneous external systems. Instead, the client enforces consistency and abstraction, so the model communicates through a uniform channel regardless of the underlying resource.  \autoref{lst:client}  illustrates how the client \emph{constructs the JSON request} (i.e., \emph{``query the student table for the top 3 by credits''}) and POSTs it to the server. This framing ensures the LLM never touches the database interface directly and that downstream components can apply uniform validation and logging.

    \item \textbf{MCP Server}: The server provides controlled access to external resources. Upon receiving a request from the client, it validates the message, enforces policy restrictions (e.g., read-only SQL), executes the operation on the outside resource, and returns results in the standardized MCP format. For instance, an MCP server may connect to a SQL database, call a \textit{REST API}, or query a proprietary dataset. By encapsulating these details, the server hides complexity from both the client and the LLM, ensuring modularity and extensibility.  \autoref{lst:server}  presents a compact handler for \texttt{/mcp} that (i) reads the \texttt{sql} parameter, (ii) executes it on the ``student'' table, and (iii) returns a standardized JSON \texttt{result}. This result then flows back through the MCP client to the LLM, which integrates the retrieved rows into its generated response and presents the final natural-language answer to the user.

\end{itemize}

Please note that the end user interacts directly with the LLM through a conversational or application interface. The user formulates a query, provides context, or issues a command. Importantly, the user does not need to know which external resources or tools may be involved. From the user’s perspective, the LLM acts as the single entry point for intelligent reasoning and task execution.

\subsection{MCP Security and Limitations}
\label{subsec:mcpsecurity} 

 MCP provides several intrinsic security properties at the protocol level. These constitute a minimal security baseline: First, MCP enforces strict compliance with a JSON-based request–response schema that is compatible with JSON-RPC. All messages must conform to predefined structural rules, which reduces the risk of arbitrary injection or malformed input.  Second, at session initialization, the MCP server must explicitly declare the tools and operations it supports, including the expected parameter structures. This handshake ensures that no undeclared or hidden functionalities can be invoked. In effect, this mechanism acts as a protocol-level whitelist, constraining interactions to a well-defined operational space.  Finally, every MCP request carries a unique identifier that must be echoed in the corresponding response. This design enforces consistent pairing between request and response, prevents message confusion or replay within a session, and provides basic traceability across the communication channel.  

 However,  the native safeguards provided by MCP are limited in scope and primarily oriented toward basic validation and traceability.  
MCP lacks intrinsic support for two fundamental security dimensions: authenticity and confidentiality of request–response exchanges. Without these guarantees, the framework cannot inherently verify the legitimacy of a message’s origin or ensure that its content has not been modified in transit. Likewise, the absence of built-in confidentiality protections leaves sensitive information exchanged between clients and servers potentially exposed to interception or leakage.  
Consequently, many MCP implementations supplement the baseline framework with additional cryptographic mechanisms such as digital signatures, message authentication codes, and secure transport protocols to provide end-to-end authenticity and confidentiality guarantees. \looseness=-1

	\section{Problem Statement and Challenges}
\label{problem_and_motivation}

\subsection{Motivation and Problem Statement} 
  
As discussed in \S\ref{subsec:mcpsecurity}, while MCP incorporates basic safeguards such as input validation, capability declaration, and request traceability, it notably lacks built-in support for two fundamental security properties: request–response authenticity and confidentiality. As a result, developers of MCP servers frequently implement custom cryptographic mechanisms to provide these guarantees, for instance by incorporating encryption, message authentication codes, or digital signatures at the application layer.

However, the reliance on developer-supplied cryptographic logic introduces a significant risk of misconfiguration and misuse. Prior studies of cryptographic libraries and APIs have consistently demonstrated that developers often select weak primitives, apply insecure modes of operation, or omit essential steps such as randomness generation and key validation. In the context of MCP, such misuses are particularly concerning: insecure implementations may allow adversaries to forge responses, tamper with sensitive requests, or exfiltrate confidential data through improperly protected channels. Given that MCP servers often act as trusted gateways to databases, proprietary APIs, and sensitive user information, the consequences of cryptographic misuse extend well beyond the protocol itself and can compromise the integrity of the broader system. Despite the prevalence of these risks, there is currently no systematic approach tailored to identifying cryptographic misuse within MCP implementations. Existing static and dynamic analysis tools are designed for general-purpose applications and lack the domain-specific awareness of MCP’s message structures, lifecycle stages, and protocol-specific interaction patterns.  
 \textit{Our goal, therefore, is to design a domain-specific analysis tool that systematically detects and characterizes cryptographic misuses in MCP implementations, thereby providing practitioners with actionable insights to harden their deployments.}

\subsection{Challenges}
Detecting cryptographic misuse in the MCP ecosystem is far from straightforward. Unlike traditional software systems, MCP servers introduce unique complexities that fundamentally hinder the applicability of existing program analysis and security auditing tools. The challenges are not limited to simple engineering obstacles such as parsing multiple languages or instrumenting code; rather, they reflect deeper issues in modeling heterogeneous implementations, reasoning about dynamic execution semantics, and interpreting the security implications of cryptographic API usage.

\vspace{2mm}
\noindent\textbf{C1. Multi-Language Heterogeneity in Implementations.}
Within the MCP ecosystem, server implementations are not confined to a single programming language. Owing to the protocol’s openness and its demand for cross-platform deployment, MCP servers are typically realized in more than ten different languages, including Python, Java, C++, PHP, Swift, and Go. This heterogeneity enhances flexibility and broadens adoption, but it simultaneously creates formidable challenges for static analysis tasks such as cryptographic misuse detection.  \looseness=-1

Cryptographic misuse detection typically involves identifying crypto-related API invocations, tracing their data-flow dependencies, and validating whether parameters conform to established security practices. Most existing tools are designed for a single language and lack the cross-language modeling capability required to analyze MCP’s heterogeneous ecosystem. The difficulty lies not only in parsing distinct syntax and grammar but also in capturing subtle semantic differences between languages. As summarized in  \autoref{tab:mcp-languages}, implementations vary widely across several dimensions: typing discipline (e.g., strongly typed Java, Swift, and Go versus dynamically typed Python, PHP, and JavaScript), memory management (manual in C++ versus automatic in managed environments), concurrency models (traditional threads, event-driven loops, or goroutines), and the prevalence of implicit defaults (e.g., padding or mode assumptions in PHP and Node.js).

 

\begin{table}[htbp]
\centering
\scriptsize
\renewcommand{\arraystretch}{1.3}
\setlength{\tabcolsep}{5pt}

\caption{Language Heterogeneity in MCP servers}
\label{tab:mcp-languages}

\begin{tabular}{l l c c c c}
\toprule
\textbf{Language} & \textbf{Library Example} & \textbf{Typing} & \textbf{Manual Memory} & \textbf{Async} & \textbf{Implicit Defaults} \\
\midrule
Java & \texttt{javax.crypto} & Strong & \xmark & \xmark & \xmark \\
Python & \texttt{PyCryptodome} & Weak & \xmark & \xmark & \cmark \\
C++ & OpenSSL, Botan & Strong & \cmark & \xmark & \xmark \\
PHP & OpenSSL extension & Weak & \xmark & \xmark & \cmark \\
Swift & CommonCrypto & Strong & \xmark & \xmark & \xmark \\
JavaScript & \texttt{crypto} module & Weak & \xmark & \cmark & \cmark \\
Go & \texttt{crypto/aes} & Strong & \xmark & \cmark & \xmark \\
\bottomrule
\end{tabular}
\end{table}

These discrepancies fundamentally affect how cryptographic APIs are invoked and how their security properties can be validated. For instance, detecting insecure use of AES/ECB mode in Java requires analyzing structured class hierarchies and checked exceptions, while the same misuse in Python is hidden within dynamically imported modules and aliased functions. Similarly, C++ demands precise reasoning over pointer semantics and API state machines, whereas PHP introduces risks through runtime configuration and implicit defaults. A single-language analyzer cannot seamlessly adapt to these variations, leaving significant blind spots in detecting misuse across the MCP ecosystem.

\vspace{2mm}
\noindent\textbf{C2. Weakly Coupled Functions and Implicit Flows.} 
The second challenge arises from the way functions are organized within individual MCP tools. Unlike conventional software modules, where functions are often explicitly linked through internal call relationships, MCP tools typically expose a set of weakly coupled functions without a predetermined orchestration order. At runtime, it is the LLM (guided by natural language prompts) that decides how to combine these functions to accomplish a task. While this plugin-style modularity provides flexibility and extensibility, it also obscures the true execution paths and creates opportunities for cryptographic misuses that are invisible to traditional static analysis.

\begin{wrapfigure}{r}{0.5\linewidth}
    \centering
    \includegraphics[width=\linewidth]{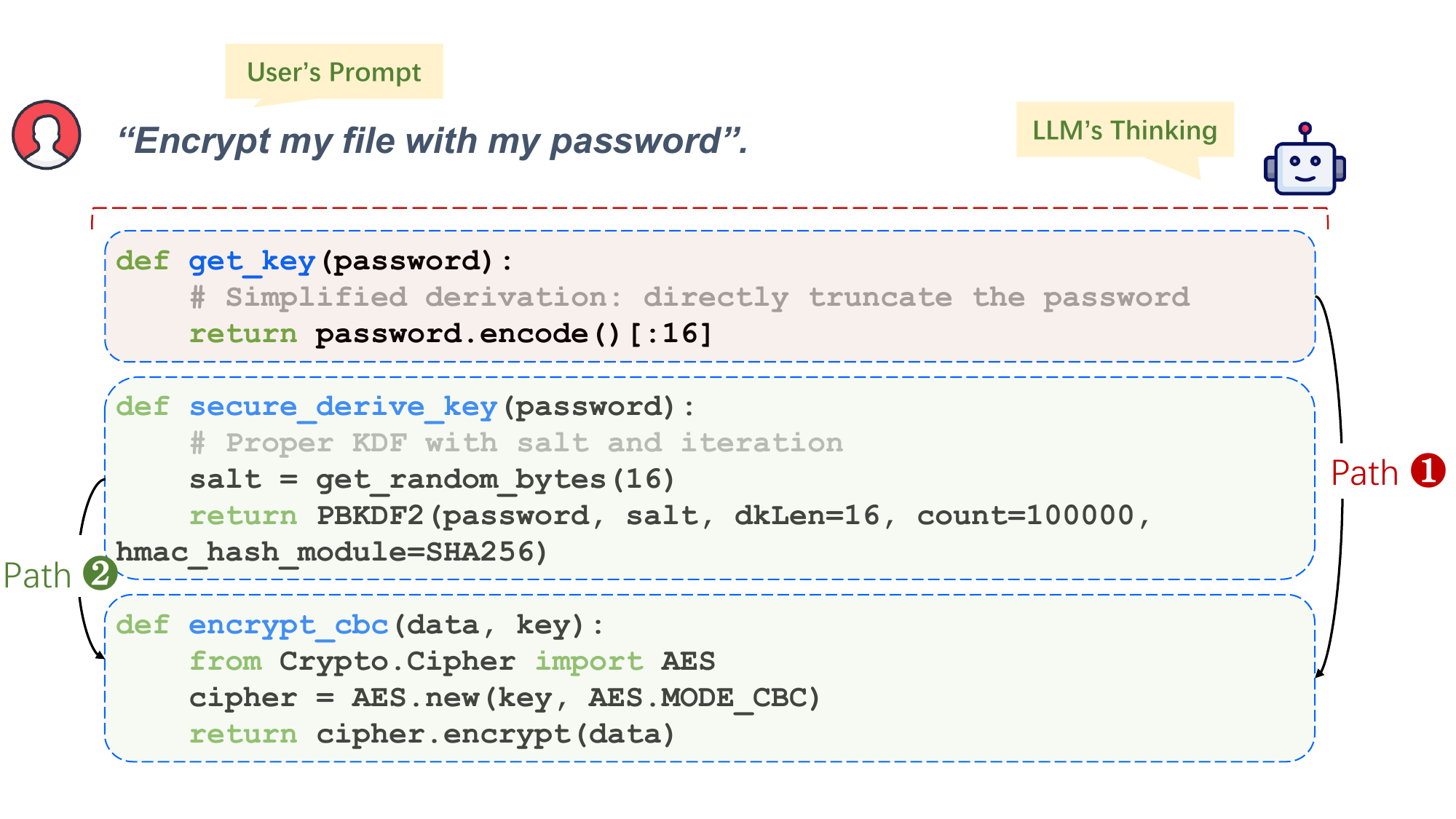}
    \caption{An MCP tool with weakly coupled functions for key derivation and encryption}
    \label{lst:derivekey}
\end{wrapfigure}

To demonstrate this challenge, we provide a simplified example shown in 
\autoref{lst:derivekey}. Individually, these functions appear legitimate: \texttt{derive\_key()} outputs a key-like string from a password, and \texttt{encrypt\_cbc()} performs AES encryption. Yet, if a user issues a prompt such as \emph{``encrypt my file with a password''}, the LLM may compose these functions by first deriving the key with \texttt{derive\_key()} and then encrypting with \texttt{encrypt\_cbc()}. The resulting execution path produces \textit{AES encryption with a weakly derived key}, which is a severe misuse. Importantly, this flaw does not originate from a single function but emerges from the implicit composition of otherwise ``reasonable'' utilities.  In contrast, the same instruction could also lead the LLM to select an alternative combination (e.g., a more secure \texttt{secure\_derive\_key()} function with a proper KDF and a \texttt{encrypt\_cbc()} function with randomized IVs), resulting in a secure execution. The critical point is that these different execution paths are not encoded as explicit control-flow edges in the tool’s code. Instead, they exist only as potential compositions resolved at runtime, outside the scope of conventional control-flow graph (CFG) construction.  

This architectural property highlights why cryptographic misuse detection in MCP cannot rely solely on traditional static analysis. Whereas classical tools reason over explicitly defined function calls, in MCP the decisive factor is the LLM’s runtime orchestration of weakly coupled functions. Misuses therefore arise not from isolated function implementations but from implicit execution paths that materialize only during prompt-driven interaction. Addressing this challenge requires analysis techniques that move beyond intra-function correctness and explicitly account for dynamic, runtime composition within tools.

\vspace{2mm}
\noindent\textbf{C3: Detecting Misuse from Static Dependencies.}
Even after achieving cross-language modeling (C1) and accurate identification of data dependencies (C2), a further challenge lies in determining whether cryptographic APIs are being misused within MCP implementations. Cryptographic misuse is not always syntactically obvious: the same API may be legitimate in one context yet insecure in another. Distinguishing misuse thus requires semantic reasoning about the purpose of an operation and the surrounding security context, not just identifying the function call itself. 


\begin{figure*}[htbp]
\centering

\begin{minipage}[t][5.2em][t]{0.47\textwidth}  
\begin{minted}[fontsize=\scriptsize, linenos, breaklines, numbersep=3pt]{python}
import hashlib
def checksum(file_path):
    with open(file_path, "rb") as f:
        data = f.read()
    return hashlib.md5(data).hexdigest()
\end{minted}
\end{minipage}
\hfill
\begin{minipage}[t][5.2em][t]{0.47\textwidth}  
\begin{minted}[fontsize=\scriptsize, linenos, breaklines, numbersep=3pt]{python}
import hashlib
def store_password(password):
    # Insecure: directly hash password with MD5
    return hashlib.md5(password.encode()).hexdigest()
\end{minted}
\end{minipage}
\caption{Contrasting uses of MD5: benign file checksums (left) vs. insecure password storage (right)}
\label{lst:md5}
\end{figure*}

A classic example is the use of MD5. In some cases, MD5 is employed merely for file integrity checking, which may be benign; in others, it is applied in security-sensitive contexts such as password storage, which constitutes a critical misuse. The contrast is illustrated in \autoref{lst:md5}. As shown in the figure, the first function computes a checksum for detecting accidental file corruption. Although MD5 is broken in terms of collision resistance, such use may be acceptable where adversarial manipulation is not a concern. The second function, however, applies MD5 for password storage, a context where its weaknesses are catastrophic: lack of salting, key stretching, and reliance on a broken hash function make it trivially vulnerable to offline brute-force attacks.  

These examples demonstrate that identifying an MD5 invocation alone is insufficient to judge security. Correct classification requires reasoning about intent (checksum vs password hashing), data type (arbitrary file vs sensitive credentials), and security requirements. In MCP, this difficulty is compounded by cross-language diversity, as the same API misuse may appear in Python (\texttt{hashlib.md5()}), Java (\texttt{MessageDigest.getInstance("MD5")}), or PHP (\texttt{openssl\_digest("md5")}).  
Thus,  the core challenge is to map observed API invocations to their security semantics and to decide whether they align with or violate best practices. Without this capability, tools risk either over-reporting benign uses or, worse, overlooking subtle yet dangerous misuses.

\section{Design of \sysname}

 We designed \sysname (\textsc{M}isuse \textsc{I}n \textsc{CRY}ptography), which is designed to act as a ``microscope'' for MCP servers. We first outline the key solutions (\S\ref{subsec:keysolution}) to the identified challenges and then present the detailed design (\S\ref{subsec:design}). \looseness=-1

 \subsection{Key Solutions}
 \label{subsec:keysolution}
\vspace{2mm}
\noindent\textbf{(S1) Cross-Language Abstraction via Unified IR.} 
The first challenge (C1) highlights that the heterogeneity of MCP server implementations across many programming languages makes it extremely difficult to build a unified analysis framework. To address this, we abstract away language-level idiosyncrasies by converting source code into abstract syntax trees (ASTs), which provide a structured yet language-agnostic view of program semantics. 
Building on ASTs, we systematically extract cryptography-related function calls, including their invocation sites, input parameters, return values, and surrounding context. For instance, we capture not only the literal arguments to a cryptographic API (e.g., key size, cipher mode) but also the symbolic variables whose values may flow into these calls. To enable cross-language analysis, we normalize all such information into a structured intermediate representation (IR) expressed in JSON. Each IR node encodes the API name, its lexical scope, parameter semantics, produced variables (such as keys, ciphertexts, or IVs), and the inferred data dependencies.\looseness=-1
 

\vspace{2mm}

\noindent\textbf{(S2) Dependencies Construction via Must/May Analysis.}  
The second challenge (C2) highlights that the plugin-style architecture of MCP tools exposes functions as weakly coupled utilities without explicit invocation chains, which makes traditional control-flow analysis (e.g., CFGs) ineffective for recovering inter-function dependencies. To overcome this limitation, we adopt a two-level dependency reconstruction strategy.  First, we perform \emph{Def–Use Analysis}, which in our context serves as a form of \emph{must analysis}. From the IR, we extract each function’s input parameters and output variables, and construct a global interprocedural Def–Use graph. This graph encodes deterministic dependencies: whenever a variable definition can be unambiguously linked to a subsequent use, a must edge is inserted. In this way, the graph recovers explicit and verifiable data flows across functions that would otherwise remain disconnected.  

\begin{wrapfigure}{r}{0.5\linewidth}
    \centering
    \includegraphics[width=\linewidth]{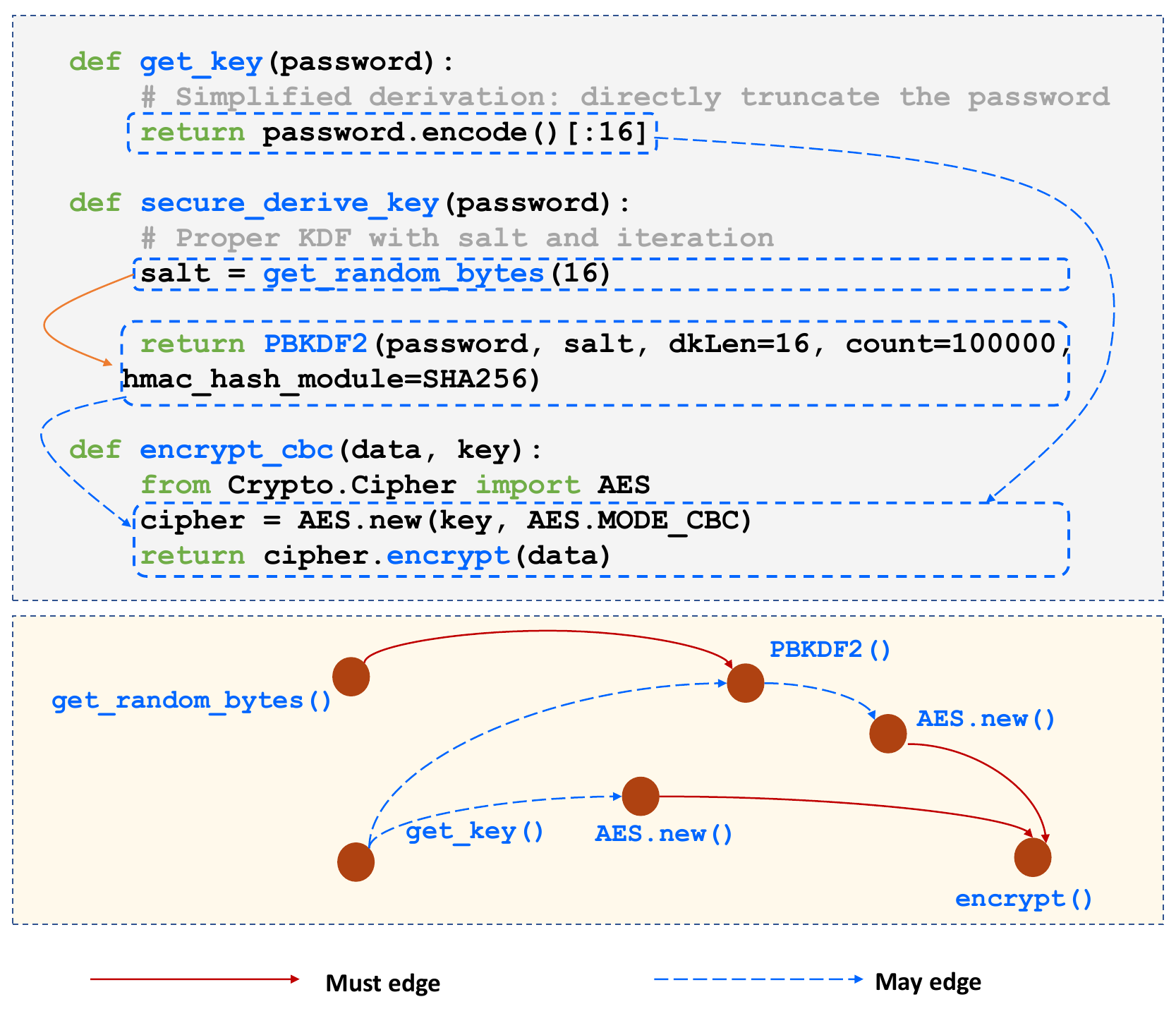}
    \caption{Must vs. May Dependency in Key Derivation}
    \label{fig:mustmay}
\end{wrapfigure}

However, MCP also involves numerous implicit dependencies that cannot be captured by must analysis alone. For example, two functions may appear independent in code, yet when orchestrated by the LLM in response to a prompt, they may operate on the same underlying resource (e.g., a file path, URL, or storage key). Such latent relationships escape Def–Use Analysis because no explicit variable passing occurs. To account for these cases, we introduce \emph{May Analysis}. Specifically, we normalize IR parameters by abstracting resource identifiers (paths, URLs, bucket/key pairs) into canonical fingerprints, and by unifying variable references into common-name entities that reflect their ultimate source. Whenever multiple functions are detected to access the same fingerprint or entity, we conservatively insert may edges to represent potential dependencies.  
By combining deterministic must edges from Def–Use Analysis with conservative may edges from resource-based normalization, we construct a comprehensive dependency graph. This hybrid graph not only recovers explicit data flows but also exposes hidden, prompt-driven dependencies between otherwise unrelated functions. As such, it provides a structured foundation for reasoning about execution order and for reliably identifying cryptographic misuse in MCP servers.

As shown in \autoref{fig:mustmay}, the use of \texttt{salt} inside \texttt{secure\_derive\_key} creates a 
\textit{must edge}, since the variable is deterministically defined by 
\texttt{get\_random\_bytes(16)} and directly consumed by \texttt{PBKDF2}. 
This represents a concrete and unambiguous dependency.  By contrast, the \texttt{key} variable introduces a \textit{may edge}: depending on runtime orchestration, it may be derived insecurely via 
\texttt{get\_key(password)} (simple truncation) or securely via 
\texttt{secure\_derive\_key(password)} (proper KDF with salt). 
Because this dependency cannot be conclusively resolved at the static level, 
it must be conservatively modeled as a potential relation. 

\vspace{2mm}
\noindent\textbf{(S3) Data Flow Tracking via Taint Analysis.}
Building on the intermediate representation (S1) and dependency graph constructed (S2), we introduce \textit{Taint Analysis}  as the core mechanism for detecting cryptographic misuse.  This approach systematically tracks the propagation of potentially untrusted data through program execution, thereby uncovering misuse patterns that would otherwise remain hidden.  Concretely, the analysis first identifies \textit{taint sources}, such as user inputs,  configuration files, or environment variables, which may inject untrusted data into the system. 
It then designates \textit{taint sinks}, including cryptographic API calls, data output operations, and file writes, as sensitive nodes where 
misuse may manifest.  Next, the taint analysis performs \textit{propagation} along the dependency graph.  
Whenever a variable obtains data from a taint source, its taint status is preserved and propagated through subsequent assignments, function calls, 
and return values. This mechanism explicitly captures the flow of untrusted inputs into sensitive cryptographic operations, while retaining contextual 
awareness across function boundaries.  Finally, the resulting propagation chains are checked against a set of 
\textit{misuse detection rules} (See \S\ref{subsec:design}). These rules encode well-known insecure patterns, such as hard-coded keys, fixed initialization vectors, 
weak algorithms (e.g., MD5), or missing authentication steps. If a propagation path matches one of these patterns, the corresponding operation is flagged as a misuse. \looseness=-1 

 \subsection{Detailed Design}
 \label{subsec:design}
\begin{figure*}
    \centering
    \includegraphics[width=1\linewidth]{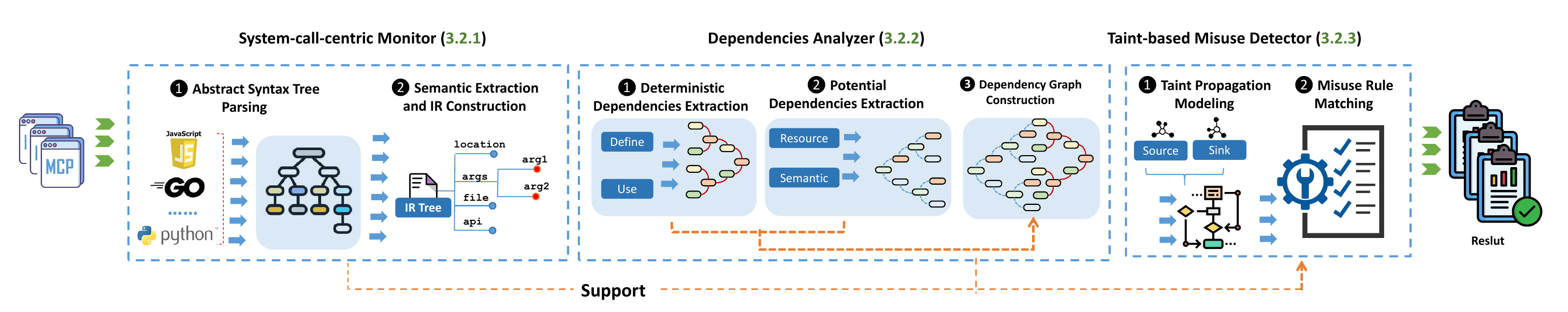}
    \caption{Design of \sysname}
    \label{fig:mcpscope}
\end{figure*}

 As shown in \autoref{fig:mcpscope}, \sysname introduces a cross-language intermediate representation to unify cryptographic API usage, a hybrid dependency analysis to recover both explicit and implicit function relationships in plugin-style architectures, and a taint-based misuse detector to trace data flows and enforce established cryptographic rules.

 \subsubsection{AST-Guided IR Extractor}
To address \textbf{C1}, the \sysname leverages the structured representation capability of ASTs to systematically parse MCP server source code and construct an IR suitable for cross-language cryptographic misuse detection.

\smallskip
\noindent\textbf{(Step I) Abstract Syntax Tree Parsing.}  We first utilize ecosystem-specific parsers (e.g., Python’s \texttt{ast}, JavaScript’s \texttt{@babel/parser}) to convert source code into ASTs. To enrich their contextual sensitivity, we enhance the raw ASTs in two key ways: First, each node is assigned a reverse reference to its parent, enabling contextual awareness (e.g., detecting whether a function call occurs on the right-hand side of an assignment). Second, a scope stack and symbol table are maintained to track variable bindings and visibility, which provides the basis for subsequent Def–Use analysis (See \S\ref{subsec:com2}).

\smallskip
\noindent\textbf{(Step II) Semantic Extraction and IR Construction.}
On top of the enriched ASTs, the system traverses all function call nodes and extracts their key features, including call name, file location, scope, and parameters. Each call is then normalized into a language-independent IR unit with the following properties: (i) All calls are uniformly encapsulated into IR nodes, allowing for consistent analysis across different languages. (ii) Parameters are semantically tagged into categories such as constant, \texttt{list\_literal}, \texttt{dict\_literal}, \texttt{function\_return}, and \texttt{variable}. 


  \subsubsection{Dependencies Analyzer}
  \label{subsec:com2}

  We leverage the IR as the carrier to recover inter-function relationships in the plugin-style MCP architecture, where explicit call chains are often absent. As shown in \autoref{alg:dep-extraction}, we design a two-level dependency extraction mechanism that combines \emph{must-dependence} and \emph{may-dependence} analysis,  followed by graph construction.

\smallskip
\noindent\textbf{(Step I) Deterministic Dependencies Extraction.} We apply a lightweight Def--Use analysis over IR nodes. Each call node specifies both its produced variable (\texttt{produced\_as}) and consumed arguments. When a variable defined in one call appears as a parameter of another call, the system matches the pair to form a deterministic Def $\rightarrow$ Use edge. For example, if  the first call generates a random \texttt{salt} and stores it in variable \texttt{salt\_val}, and  the second call later passes \texttt{salt\_val} as an argument to a key derivation API (e.g., \texttt{PBKDF2}), the framework records a must-dependence edge. By exhaustively scanning all call pairs within the file scope, the system constructs the complete set of explicit dependencies.

\begin{algorithm}[t]
\scriptsize
\caption{\footnotesize Dependency Extraction}
\label{alg:dep-extraction}
\KwIn{IR set $\mathcal{I}$ of all call nodes from project $P$}
\KwOut{Dependency graph $G=(V,E)$ with must- and may-dependence edges}

Initialize $V \gets \mathcal{I}$, $E \gets \emptyset$\;

\BlankLine
\textbf{Step 1: Must-dependence extraction (Def--Use)} \;
\ForEach{call node $c \in \mathcal{I}$}{
    \ForEach{argument $a \in c.arguments$}{
        \If{$a.type = variable$}{
            $v \gets a.value$\;
            \ForEach{call node $d \in \mathcal{I}$}{
                \If{$d.produced\_as = v$}{
                    add edge $(d \rightarrow c)$ to $E_{must}$\;
                }
            }
        }
    }
}

\BlankLine
\textbf{Step 2: May-dependence extraction} \;
\ForEach{call node $c \in \mathcal{I}$}{
    extract $fp(c)$ from arguments of type \texttt{constant} (path, topic, ID)\;
    assign semantic label $sem(c)$ from API categorization\;
}
\ForEach{pair $(c_i, c_j)$ with $i \neq j$}{
    \If{$(sem(c_i), sem(c_j)) \in$ risky patterns}{
        \If{$fp(c_i) = fp(c_j)$ \textbf{or} shared variable in parameters}{
            add edge $(c_i \leadsto c_j)$ to $E_{may}$\;
        }
    }
}

\BlankLine
\textbf{Step 3: Graph construction}\;
$E \gets E_{must} \cup E_{may}$\;
\Return $G=(V,E)$\;

\end{algorithm}

\smallskip
\noindent\textbf{(Step II)  Potential Dependencies Extraction.} Explicit Def--Use relationships, however, are insufficient for weakly coupled modules common in MCP. To capture implicit yet semantically meaningful relationships, we introduce may-dependence edges through  {resource fingerprinting} and {API semantic categorization}: \looseness=-1  
\begin{itemize}
    \item \textbf{Resource fingerprinting} extracts path names, message topics, or identifiers from arguments of type \texttt{constant}, treating them as resource anchors across calls.  For example, a \texttt{write\_file} call that outputs data to 
\texttt{"user.db"} and a subsequent \texttt{read\_file} call accessing 
the same path are linked via the constant \texttt{"user.db"} as a 
resource fingerprint, forming a may-dependence edge.

\item \textbf{Semantic categorization} groups APIs into high-level functions such as 
\texttt{protect}, \texttt{upload}, or \texttt{mask}, enabling consistent analysis 
across languages. A may-dependence edge is introduced when two calls both operate 
on the same resource (via a shared fingerprint such as a filename or overlapping variable names such as ``\textit{key}'') and 
their functions match a high-risk pattern. 
For example, consider a call 
\texttt{encrypt(data, key)} categorized as \texttt{protect}, followed by a call 
\texttt{upload("user.db")} categorized as \texttt{upload}. Although no variable 
is explicitly passed between them, both share the fingerprint \texttt{"user.db"}. 
The system therefore records a may-dependence edge \texttt{encrypt} $\rightarrow$ 
\texttt{upload}, modeling the implicit flow of sensitive data from local 
protection to external exposure.
\end{itemize}

\smallskip
\noindent\textbf{(Step III) Dependency Graph Construction.}
Finally, both must- and may-dependence edges are integrated into a unified dependency graph. Each IR node is represented as a graph vertex, while edges denote either must or may relationships. The resulting graph is directed, typed, and layered: nodes encode API calls with their attributes, edges encode dependence semantics, and subgraphs naturally emerge for modules or plugins. This enriched dependency graph not only preserves precision from deterministic analysis but also broadens coverage by incorporating semantic cues, forming the structural foundation for downstream misuse detection.

\subsubsection{Taint-based Misuse Detector}

   In the third component, we apply taint analysis to uncover insecure configurations and data leakage paths in cryptographic API usage. The process consists of two steps:   taint propagation modeling and misuse rule matching.

\begin{table}[htbp]
\centering
\scriptsize
\renewcommand{\arraystretch}{1.05}
\setlength{\tabcolsep}{2pt}

\caption{Eight Rules for Cryptographic API Misuse Detection}
\label{tab:rules}

\begin{tabular}{p{0.8cm}p{2.8cm}p{9.5cm}} 
\toprule
\textbf{ID} & \textbf{Rule} & \textbf{Description} \\
\midrule
R1 & Fixed Key / API Key & Hard-coded encryption or API keys directly embedded in code, making them easily recoverable and reusable by attackers~\cite{chen2024towards,ami2022crypto,sun2023cryptoeval,wang2024cryptody,piccolboni2021crylogger,mandal2024belt}. \\
\midrule
R2 & Fixed IV / Salt & Use of constant IVs or salts, which weaken randomness and compromise security~\cite{chen2024towards,ami2022crypto,sun2023cryptoeval,wang2024cryptody,piccolboni2021crylogger}. \\
\midrule
R3 & Weak Hash Functions & Usage of MD5, SHA1, or other broken hash algorithms in security-sensitive contexts~\cite{chen2024towards,ami2022crypto,torres2023runtime,xia2025beyond,piccolboni2021crylogger}. \\
\midrule
R4 & Insecure Key Derivation Configuration & Use of password-based encryption (PBE) with insufficient iterations or weak parameter settings~\cite{chen2024towards,ami2022crypto,mandal2024belt}. \\
\midrule
R5 & Static Seed in PRNG & PRNG initialized with a fixed seed, producing deterministic random sequences~\cite{chen2024towards,ami2022crypto,sun2023cryptoeval,piccolboni2021crylogger}. \\
\midrule
R6 & ECB Mode Usage & Use of ECB block cipher mode, which leaks plaintext structure through repeated patterns~\cite{chen2024towards,ami2022crypto,wang2024cryptody,piccolboni2021crylogger,mandal2024belt}. \\
\midrule
R7 & Missing Integrity Protection & Encryption applied without authentication (e.g., AES without MAC/GCM), allowing undetected tampering~\cite{chen2024towards,ami2022crypto}. \\
\midrule
R8 & Deprecated Alg/APIs & Use of outdated cryptographic primitives (e.g., DES, RC4) or unsafe APIs~\cite{chen2024towards,ami2022crypto,piccolboni2021crylogger}. \\
\bottomrule
\end{tabular}
\end{table}

\smallskip
\noindent\textbf{(Step I) Taint Propagation Modeling.}
Over the constructed dependency graph, each argument and produced variable is semantically annotated (e.g., \texttt{constant}, \texttt{variable}, \texttt{function\_return}, \texttt{dict\_literal}, \texttt{list\_literal}).  
By recursively tracing the Def--Use chains, the system resolves intermediate variables to their concrete sources (e.g., literals or fixed configurations), thereby exposing hidden hard-coded secrets or unsafe defaults.  
Taint analysis is then performed: external inputs are defined as \emph{sources}, while operations such as printing, persistence, or network transmission are defined as \emph{sinks}. A forward traversal of the graph is conducted to determine not only whether a flow exists but also \emph{how it flows}. To systematically capture common pitfalls, we summarize ten representative rules of cryptographic API misuse in ~\autoref{tab:rules}. Specifically, the analysis evaluates whether a propagated key is hard-coded, whether an IV is constant, whether a weak hash function is used, or whether randomness is seeded deterministically.  The analysis spans multiple functions and files while maintaining consistent interprocedural context.

\smallskip
\noindent\textbf{(Step II) Misuse Rule Matching.}
Once taint propagation identifies insecure propagation chains, the system applies misuse detection rules to flag concrete vulnerabilities. A project is marked as misusing cryptography if a tainted artifact reaches a sink under insecure conditions, such as constant keys, fixed IVs, weak hash algorithms, or predictable random seeds. The combination of dependency graph analysis and taint semantics thus enables systematic and fine-grained detection of cryptographic misuses across heterogeneous MCP implementations. \looseness=-1
	\section{Evaluation}

\subsection{Experiment Setup}
\label{der_evaluation}

\paragraph{Data Collection.}
We systematically collected MCP server projects from GitHub API and multiple external registries, including \textit{Smithery, Pulse MCP, Cursor Directory, Awesome MCP, Glama AI, \textit{Mcpmarket}}, and \textit{Modelcontextprotocol.io.} A customized crawler was implemented to automatically retrieve both the source code and the corresponding metadata of MCP servers.  In total, we obtained 9,403 MCP servers, covering a wide range of functional categories. Among them, Developer Tools dominate the dataset, with more than 2,300 instances from \textit{Mcpmarket} alone.   To mitigate class imbalance and avoid bias from sparsely distributed categories, we grouped the remaining servers into an ``\textit{Other}'' category, ensuring a comprehensive and representative dataset. In addition, a small portion of servers is labeled ``\textit{Unknown}'', reflecting incomplete or inconsistent metadata in public registries. 

\begin{wrapfigure}{r}{0.5\linewidth}
  \centering
  \includegraphics[width=\linewidth]{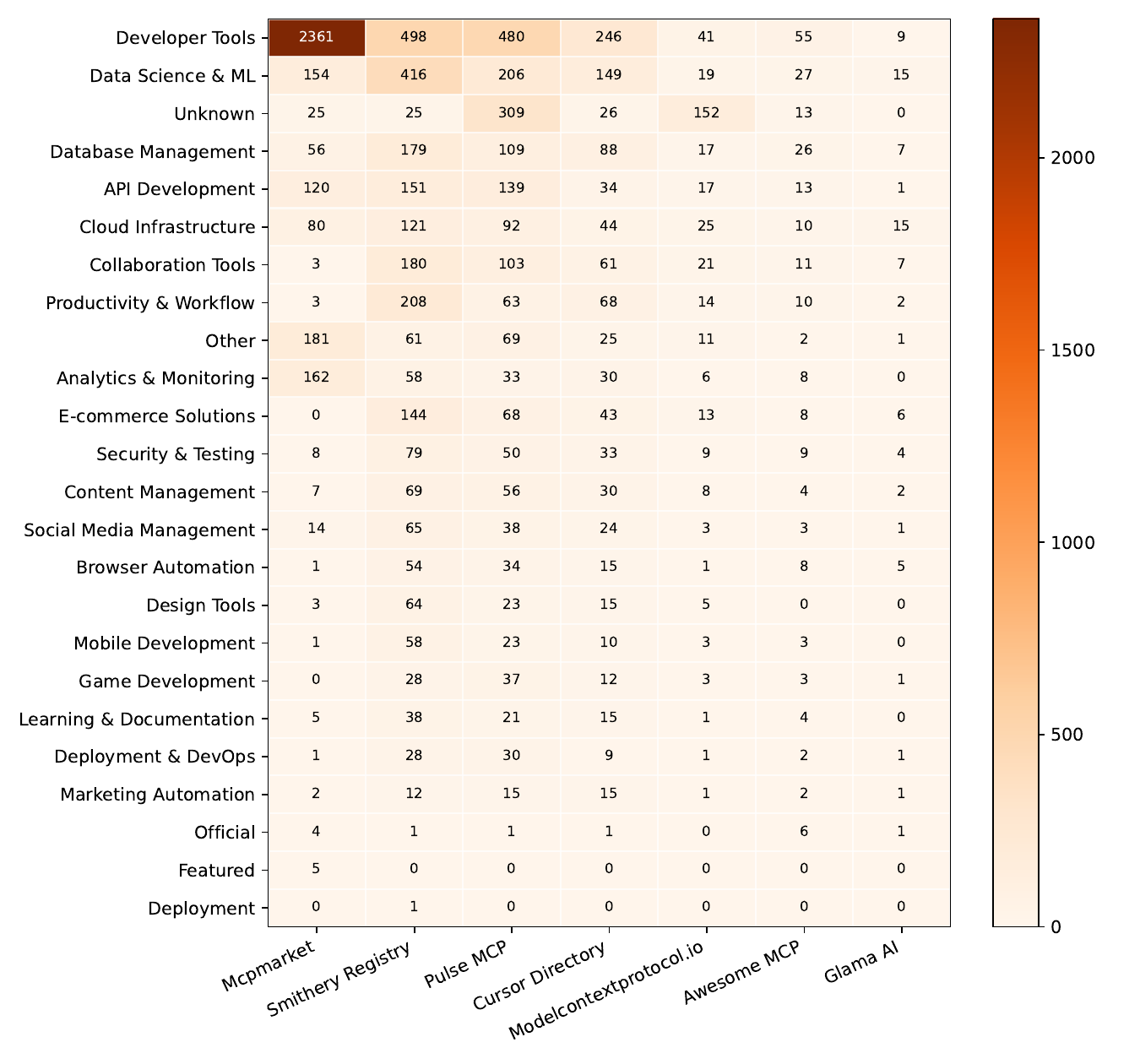}
  \caption{Distribution of MCP Servers}
  \label{fig:MCP_data_dis}
\end{wrapfigure}

As shown in \autoref{fig:MCP_data_dis}, our analysis of those servers reveals several noteworthy patterns in markets and categories distribution. First, the ecosystem demonstrates a concentrated yet multipolar structure: while Mcpmarket alone accounts for roughly one quarter of all servers, Smithery Registry and Pulse MCP also contribute substantial shares, together forming a small set of dominant platforms. Second, in terms of functionality, Developer Tools overwhelmingly dominate the landscape, with more than half of all servers ($\sim$5,000) falling into this category, underscoring the developer-centric nature of the MCP ecosystem. Finally, although secondary categories such as Data Science \& ML,  Database Management, and Web Scraping are represented, application-oriented domains such as e-commerce, social media, and content management remain marginal, suggesting that MCP adoption beyond developer-focused use cases is still in its early stages.

\vspace{2mm}
\paragraph{Execution Environment.} All experiments were conducted on a workstation equipped with an Intel(R) Core(TM) Ultra 7 258V @ 2.20 GHz processor, 32 GB RAM, and Windows 11 Pro (Version 24H2, OS Build 26100.4946). The system runs on a 64-bit x64-based architecture. This configuration offers sufficient computational power for large-scale parsing, IR construction, and cryptographic misuse detection.

\vspace{2mm}
\paragraph{Implementation.} Our analysis pipeline was primarily implemented in Python (3.11) with approximately 6,000 lines of code. Since extracting ASTs required handling language-specific features, we employed parsers in multiple languages (e.g., JavaScript and TypeScript for front-end plugins, Python for backend servers) to generate ASTs. 
\subsection{Performance of \sysname}
\begin{wrapfigure}{r}{0.45\linewidth}
  \centering
  \includegraphics[width=\linewidth]{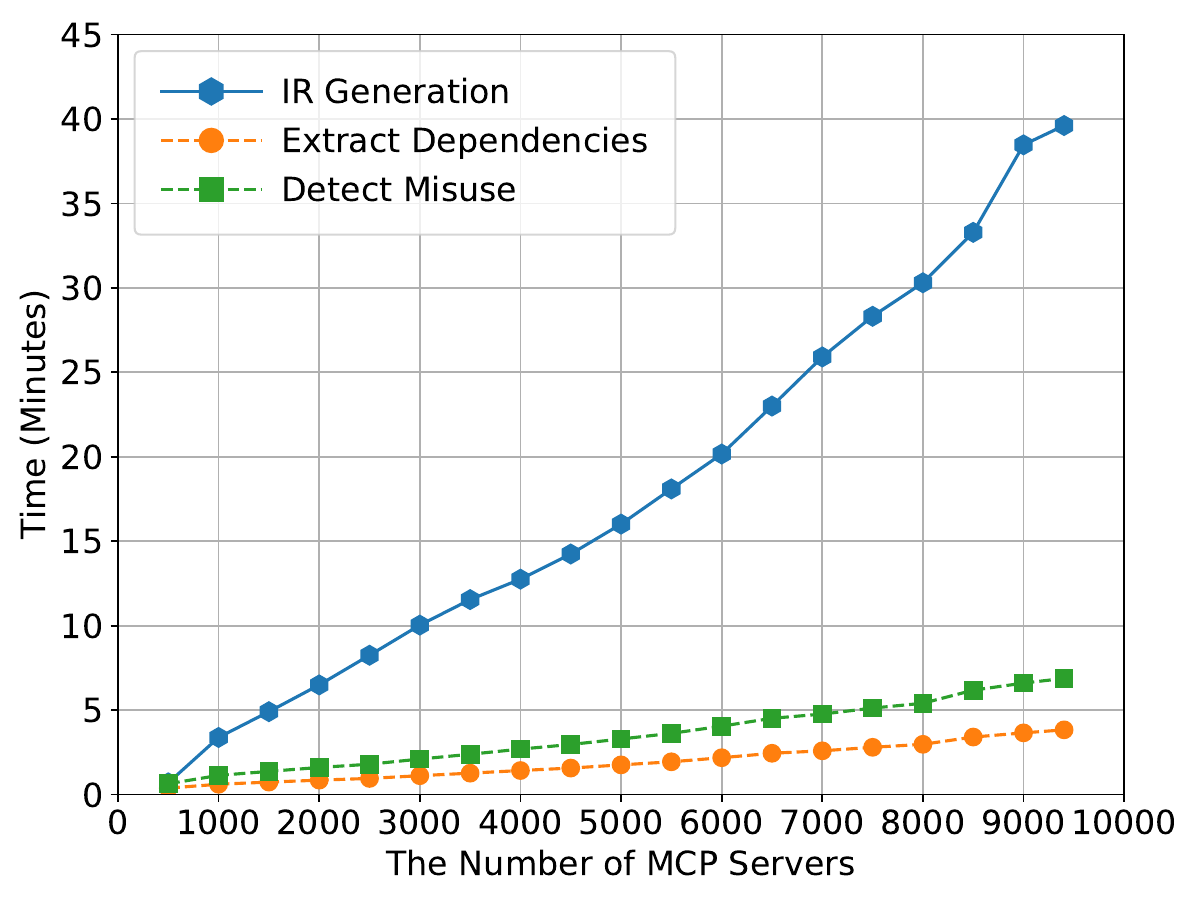}
  \caption{Performance of our \sysname}
  \label{img:MCP_tool_time}
\end{wrapfigure}

\vspace{1mm}
\paragraph{Time Overhead.} We evaluate the end-to-end performance of \sysname when applied to the entire dataset of 9,403 MCP servers. As shown in \autoref{img:MCP_tool_time}, the total execution time was broken down into three major stages: IR generation, dependency extraction, and misuse detection. The overall runtime is dominated by IR generation, which accounts for more than half of the total analysis time. This is expected, as multi-language AST parsing and normalization into the unified IR format is the most resource-intensive step. In contrast, dependency extraction (must-/may-dependence analysis) and misuse detection (taint propagation and rule matching) together contribute a smaller fraction of the total runtime, showing that the heavy lifting lies in IR construction. The distribution suggests no sharp outliers: runtime grows smoothly with dataset size, and the full corpus can be processed in a feasible timeframe on a single workstation. This indicates that \sysname is capable of handling large-scale MCP ecosystems and can be readily extended to even larger registries as they emerge.

\vspace{1mm}
\paragraph{Accuracy.} 
\sysname demonstrates high accuracy. In manual validation, 5 of 100 randomly sampled servers classified as misuse-free were found to contain misuses (5\% FN), while all 142 flagged cases were confirmed as true misuses (0\% FP). These cases largely arise from two limitations. First, \sysname relies on static IR construction and taint propagation. When cryptographic parameters are generated dynamically at runtime (e.g., IVs derived from system time or environment randomness), they may remain as unresolved variables in the IR rather than being reduced to insecure constants, causing misuses to be overlooked. Second, some projects embed cryptographic logic indirectly through wrappers, utility functions, or domain-specific libraries. If such patterns fall outside \sysname’s rule set, they may evade detection. One representative case is the \textit{MongoDB\_Atlas} server (\autoref{fig:fn-md5}), where a custom \texttt{md5} wrapper internally invoked the standard library. Because sensitive values such as API keys, realms, and nonces were concatenated into compound strings before hashing, their roles were obscured, and the weak MD5 usage was missed. This combination of API hiding and string manipulation demonstrates how indirect or obfuscated implementations can reduce static visibility and lead to false negatives.

\begin{figure}[htbp]
\centering
\begin{minted}[fontsize=\scriptsize,breaklines, numbersep=3pt]{javascript}
  private md5(data: string): string {
    return crypto.createHash('md5').update(data).digest('hex');} 
  const ha1 = this.md5(`${this.apiKey}:${authDetails.realm}:${this.privateKey}`);
  const ha2 = this.md5(`${method}:${new URL(url).pathname}`);
  const response = this.md5(`${ha1}:${authDetails.nonce}:${nc}:${cnonce}:${authDetails.qop}:${ha2}`);
\end{minted}
\caption{False negative example where \sysname missed weak MD5 usage due to custom wrapper and string concatenation}
\label{fig:fn-md5}
\end{figure}

\begin{table*}[!htb]
\centering
\caption{Language Distribution of Crypto Adoption and Misuse Across MCP Markets}

{
\setlength{\tabcolsep}{1.6pt}
\renewcommand{\arraystretch}{1.05}
\begingroup
\tiny

\begin{tabular*}{\textwidth}{@{\extracolsep{\fill}}l*{7}{@{}lllllllllllllllll@{\hspace{0.6pt}}}}
\toprule
\multirow{3}{*}{\textbf{Language}} &
\multicolumn{4}{c}{\textbf{Mcpmarket}} &
\multicolumn{4}{c}{\textbf{Smithery Registry}} &
\multicolumn{4}{c}{\textbf{Pulse MCP}} &
\multicolumn{4}{c}{\textbf{Cursor Directory}} &
\multicolumn{4}{c}{\textbf{MCP.io}} &
\multicolumn{4}{c}{\makecell{\textbf{Awesome MCP}}} &
\multicolumn{4}{c}{\textbf{Glama AI}} \\
\cmidrule(lr){2-5}\cmidrule(lr){6-9}\cmidrule(lr){10-13}
\cmidrule(lr){14-17}\cmidrule(lr){18-21}\cmidrule(lr){22-25}\cmidrule(lr){26-29}
& \multicolumn{2}{c}{Crypto} & \multicolumn{2}{c}{Misuse}
& \multicolumn{2}{c}{Crypto} & \multicolumn{2}{c}{Misuse}
& \multicolumn{2}{c}{Crypto} & \multicolumn{2}{c}{Misuse}
& \multicolumn{2}{c}{Crypto} & \multicolumn{2}{c}{Misuse}
& \multicolumn{2}{c}{Crypto} & \multicolumn{2}{c}{Misuse}
& \multicolumn{2}{c}{Crypto} & \multicolumn{2}{c}{Misuse}
& \multicolumn{2}{c}{Crypto} & \multicolumn{2}{c}{Misuse} \\
\cmidrule(lr){2-3}\cmidrule(lr){4-5}\cmidrule(lr){6-7}\cmidrule(lr){8-9}
\cmidrule(lr){10-11}\cmidrule(lr){12-13}
\cmidrule(lr){14-15}\cmidrule(lr){16-17}
\cmidrule(lr){18-19}\cmidrule(lr){20-21}
\cmidrule(lr){22-23}\cmidrule(lr){24-25}
\cmidrule(lr){26-27}\cmidrule(lr){28-29}
& \textbf{\cmark} & \textbf{\xmark} & \textbf{\cmark} & \textbf{\xmark}
& \textbf{\cmark} & \textbf{\xmark} & \textbf{\cmark} & \textbf{\xmark}
& \textbf{\cmark} & \textbf{\xmark} & \textbf{\cmark} & \textbf{\xmark}
& \textbf{\cmark} & \textbf{\xmark} & \textbf{\cmark} & \textbf{\xmark}
& \textbf{\cmark} & \textbf{\xmark} & \textbf{\cmark} & \textbf{\xmark}
& \textbf{\cmark} & \textbf{\xmark} & \textbf{\cmark} & \textbf{\xmark}
& \textbf{\cmark} & \textbf{\xmark} & \textbf{\cmark} & \textbf{\xmark} \\
\midrule
Javascript & 112 & 1555 & 17 & 1650 & 142 & 1390 & 24 & 1508 & 143 & 1614 & 12 & 1745 & 30 & 535 & 3 & 562 & 31 & 287 & 5 & 313 & 14 & 125 & 1 & 138 & 2 & 46 & - & 48 \\
Python     & 89  & 1152 & 30 & 1211 & 91  & 813  & 32 & 872  & -   & -    & -  & -    & 18 & 355 & 4 & 369 & -  & -   & - & -   & 3  & 69  & - & 72  & 1 & 30 & - & 31 \\
Go         & 11  & 131  & 4  & 138  & 6   & 53   & 3  & 56   & 5   & 108  & 1  & 112  & -  & 22  & - & 22  & 4  & 26  & 1 & 29  & -  & 11  & - & 11  & - & -  & - & - \\
Java       & 2   & 47   & 2  & 47   & 4   & 24   & -  & 28   & 1   & 38   & 1  & 38   & -  & 12  & - & 12  & -  & 8   & - & 8   & -  & -   & - & -   & - & -  & - & - \\
Rust       & -   & 37   & -  & 37   & 1   & 6    & 1  & 6    & -   & 38   & -  & 38   & -  & 6   & - & 6   & -  & 10  & - & 10  & -  & 1   & - & 1   & - & -  & - & - \\
TypeScript & 2   & 13   & -  & 15   & -   & -    & -  & -    & 2   & 23   & -  & 25   & 3  & 6   & 1 & 8   & -  & 4   & - & 4   & -  & 1   & - & 1   & - & -  & - & - \\
C\#        & -   & 26   & -  & 26   & -   & 5    & -  & 5    & -   & 9    & -  & 9    & -  & 2   & - & 2   & -  & -   & - & -   & -  & 2   & - & 2   & - & -  & - & - \\
Swift      & -   & 9    & -  & 9    & -   & -    & -  & -    & -   & 9    & -  & 9    & -  & 3   & - & 3   & -  & 1   & - & 1   & -  & 1   & - & 1   & - & -  & - & - \\
Ruby       & -   & 7    & -  & 7    & -   & 1    & -  & 1    & -   & 5    & -  & 5    & -  & -   & - & -   & -  & -   & - & -   & -  & -   & - & -   & - & -  & - & - \\
PHP        & -   & 3    & -  & 3    & 1   & 1    & -  & 2    & 2   & 2    & -  & 4    & -  & 1   & - & 1   & -  & -   & - & -   & -  & -   & - & -   & - & -  & - & - \\
\bottomrule
\end{tabular*}

\endgroup
}
\label{tab:lang_market_7markets_crypto_misuse}
\end{table*}

\subsection{Empirical Results}
In total, we identified 720 MCP (out of 9,403) servers that contained cryptographic operations, among which 142 instances (19.7\%) exhibited cryptographic misuses. To better understand the distribution and characteristics of these misuses, we conduct a detailed analysis from three complementary perspectives: (i) the distribution across different markets, (ii) the programming languages employed, (iii) the prevalence within functional categories, and (iv) the breakdown with respect to specific misuse rules.

\begin{wrapfigure}{r}{0.5\linewidth}
  \centering
  \includegraphics[width=\linewidth]{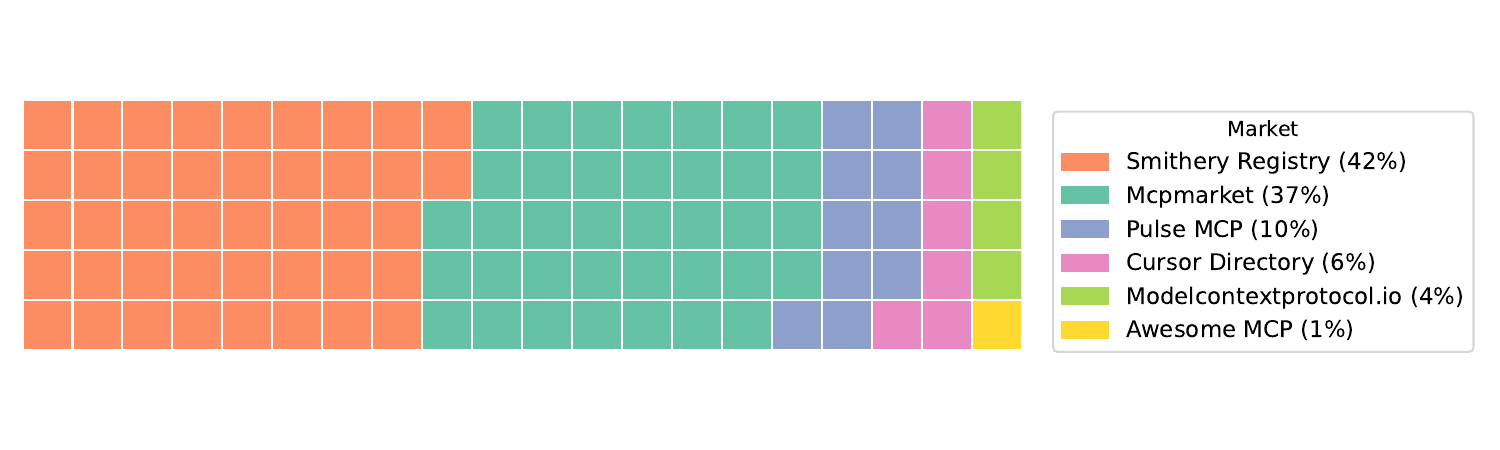}
  \caption{MCP Servers with Crypto Misuse by Market}
  \label{img:MCP_misuse_by_market}
\end{wrapfigure}

\vspace{2mm}
\noindent\textbf{(I) Market-Level Analysis of Misuse.}
We begin by examining the market-level distribution of misuses across the seven major MCP registries. It can be observed from \autoref{img:MCP_misuse_by_market} that while Mcpmarket dominates the ecosystem overall (3,196 servers, more than double the next-largest Smithery Registry with 2,538), its share of misuses (37\%) is slightly lower than Smithery’s (42\%). This indicates that Smithery Registry, despite being smaller in absolute size, harbors a disproportionately high number of insecure implementations. Pulse MCP, with 1,999 servers, contributes 10\% of the misuses, while Cursor Directory and Modelcontextprotocol.io show smaller absolute counts, but their relative misuse rates are not negligible given their more modest project bases. Finally, Awesome MCP account for only 1\% of misuses, yet their presence highlights that even niche or community-driven platforms are not immune to insecure cryptographic practices. Although the ecosystem is highly centralized around Mcpmarket, the greatest density of misuses lies in Smithery Registry, suggesting that misuse is influenced not just by market size, but also by the curation standards and developer practices characteristic of each registry.

\begin{table*}[!htb]
\centering
\caption{Cross-Market Category Distribution with Crypto Adoption and Misuse}
{
\setlength{\tabcolsep}{1pt}
 
\begingroup
\tiny
\resizebox{\textwidth}{!}{%
\begin{tabular*}{\textwidth}{@{\extracolsep{\fill}}l*{7}{@{}llllllllllllllllllllll@{}}}
\toprule
\multirow{3}{*}{\textbf{Language}} &
\multicolumn{4}{c}{\textbf{Mcpmarket}} &
\multicolumn{4}{c}{\textbf{Smithery Registry}} &
\multicolumn{4}{c}{\textbf{Pulse MCP}} &
\multicolumn{4}{c}{\textbf{Cursor Directory}} &
\multicolumn{4}{c}{\textbf{MCP.io}} &
\multicolumn{4}{c}{\textbf{Awesome MCP}} &
\multicolumn{4}{c}{\textbf{Glama AI}} \\ 
\cmidrule(lr){2-5}\cmidrule(lr){6-9}\cmidrule(lr){10-13}
\cmidrule(lr){14-17}\cmidrule(lr){18-21}\cmidrule(lr){22-25}\cmidrule(lr){26-29}
& \multicolumn{2}{c}{\textbf{Crypto}} & \multicolumn{2}{c}{\textbf{Misuse}}
& \multicolumn{2}{c}{\textbf{Crypto}} & \multicolumn{2}{c}{\textbf{Misuse}}
& \multicolumn{2}{c}{\textbf{Crypto}} & \multicolumn{2}{c}{\textbf{Misuse}}
& \multicolumn{2}{c}{\textbf{Crypto}} & \multicolumn{2}{c}{\textbf{Misuse}}
& \multicolumn{2}{c}{\textbf{Crypto}} & \multicolumn{2}{c}{\textbf{Misuse}}
& \multicolumn{2}{c}{\textbf{Crypto}} & \multicolumn{2}{c}{\textbf{Misuse}}
& \multicolumn{2}{c}{\textbf{Crypto}} & \multicolumn{2}{c}{\textbf{Misuse}} \\ 
\cmidrule(lr){2-3}\cmidrule(lr){4-5}\cmidrule(lr){6-7}\cmidrule(lr){8-9}
\cmidrule(lr){10-11}\cmidrule(lr){12-13}
\cmidrule(lr){14-15}\cmidrule(lr){16-17}
\cmidrule(lr){18-19}\cmidrule(lr){20-21}
\cmidrule(lr){22-23}\cmidrule(lr){24-25}
\cmidrule(lr){26-27}\cmidrule(lr){28-29}
& \textbf{\cmark} & \textbf{\xmark} & \textbf{\cmark} & \textbf{\xmark}
& \textbf{\cmark} & \textbf{\xmark} & \textbf{\cmark} & \textbf{\xmark}
& \textbf{\cmark} & \textbf{\xmark} & \textbf{\cmark} & \textbf{\xmark}
& \textbf{\cmark} & \textbf{\xmark} & \textbf{\cmark} & \textbf{\xmark}
& \textbf{\cmark} & \textbf{\xmark} & \textbf{\cmark} & \textbf{\xmark}
& \textbf{\cmark} & \textbf{\xmark} & \textbf{\cmark} & \textbf{\xmark}
& \textbf{\cmark} & \textbf{\xmark} & \textbf{\cmark} & \textbf{\xmark} \\ 
\midrule
Developer Tools & 161 & 2200 & 38 & 2323 & 37 & 461 & 3 & 495 & 31 & 449 & 1 & 479 & 10 & 236 & 1 & 245 & 3 & 38 & 1 & 40 & 1 & 54 & - & 55 & 1 & 8 & - & 9 \\
Data Science \& ML & 6 & 148 & 3 & 151 & 47 & 369 & 17 & 399 & 16 & 190 & 2 & 204 & 8 & 141 & 2 & 147 & 2 & 17 & - & 19 & 1 & 26 & - & 27 & - & 15 & - & 15 \\
Database Management & 4 & 52 & - & 56 & 16 & 163 & 1 & 178 & 6 & 103 & - & 109 & 6 & 82 & 1 & 87 & 1 & 16 & - & 17 & 4 & 22 & - & 26 & - & 7 & - & 7 \\
Collaboration Tools & 1 & 2 & - & 3 & 14 & 166 & 5 & 175 & 7 & 96 & 2 & 101 & 1 & 60 & 1 & 60 & - & 21 & - & 21 & - & 11 & - & 11 & - & 7 & - & 7 \\
Cloud Infrastructure & 5 & 75 & - & 80 & 15 & 106 & 3 & 118 & 7 & 85 & - & 92 & 7 & 37 & 1 & 43 & 5 & 20 & 2 & 23 & 2 & 8 & - & 10 & 1 & 14 & - & 15 \\
Productivity \& Workflow & - & 3 & - & 3 & 26 & 182 & 9 & 199 & 7 & 56 & - & 63 & 4 & 64 & 1 & 67 & - & 14 & - & 14 & 3 & 7 & 1 & 9 & - & 2 & - & 2 \\
Analytics \& Monitoring & 8 & 154 & 3 & 159 & 2 & 56 & - & 58 & 5 & 28 & - & 33 & - & 30 & - & 30 & 1 & 5 & - & 6 & - & 8 & - & 8 & - & - & - & - \\
E-commerce Solutions & - & - & - & - & 14 & 130 & 5 & 139 & 5 & 63 & - & 68 & 2 & 41 & - & 43 & - & 13 & - & 13 & 1 & 7 & - & 8 & 1 & 5 & - & 6 \\
Security \& Testing & 3 & 5 & 1 & 7 & 11 & 68 & 3 & 76 & 5 & 45 & 2 & 48 & 2 & 31 & - & 33 & - & 9 & - & 9 & - & 9 & - & 9 & - & 4 & - & 4 \\
Content Management & 1 & 6 & 1 & 6 & 10 & 59 & 3 & 66 & 3 & 53 & 1 & 55 & 2 & 28 & - & 30 & - & 8 & - & 8 & - & 4 & - & 4 & - & 2 & - & 2 \\
Social Media Management & 1 & 13 & - & 14 & 2 & 63 & - & 65 & 7 & 31 & - & 38 & 2 & 22 & 1 & 23 & 1 & 2 & - & 3 & - & 3 & - & 3 & - & 1 & - & 1 \\
Browser Automation & - & 1 & - & 1 & 9 & 45 & - & 54 & 5 & 29 & - & 34 & 2 & 13 & - & 15 & - & 1 & - & 1 & 1 & 7 & - & 8 & - & 5 & - & 5 \\
Design Tools & - & 3 & - & 3 & 5 & 59 & 2 & 62 & - & 23 & - & 23 & - & 15 & - & 15 & - & 5 & - & 5 & - & - & - & - & - & - & - & - \\
Mobile Development & - & 1 & - & 1 & 7 & 51 & 1 & 57 & 1 & 22 & - & 23 & - & 10 & - & 10 & - & 3 & - & 3 & - & 3 & - & 3 & - & - & - & - \\
Game Development & - & - & - & - & 1 & 27 & - & 28 & 2 & 35 & - & 37 & - & 12 & - & 12 & - & 3 & - & 3 & - & 3 & - & 3 & - & 1 & - & 1 \\
Learning \& Documentation & - & 5 & - & 5 & 2 & 36 & 1 & 37 & 2 & 19 & - & 21 & - & 15 & - & 15 & - & 1 & - & 1 & - & 4 & - & 4 & - & - & - & - \\
Deployment \& DevOps & - & 1 & - & 1 & 3 & 25 & - & 28 & - & 30 & - & 30 & - & 9 & - & 9 & - & 1 & - & 1 & - & 2 & - & 2 & - & 1 & - & 1 \\
Marketing Automation & - & 2 & - & 2 & 1 & 11 & - & 12 & 3 & 12 & - & 15 & 2 & 13 & - & 15 & 1 & - & - & 1 & - & 2 & - & 2 & - & 1 & - & 1 \\
API Development & 10 & 110 & 1 & 119 & 11 & 140 & - & 151 & 12 & 127 & - & 139 & 2 & 32 & - & 34 & - & 17 & - & 17 & - & 13 & - & 13 & - & 1 & - & 1 \\
Official & - & 4 & - & 4 & - & 1 & - & 1 & - & 1 & - & 1 & - & 1 & - & 1 & - & - & - & - & - & 6 & - & 6 & - & 1 & - & 1 \\
Featured & - & 5 & 1 & 4 & - & - & - & - & - & - & - & - & - & - & - & - & - & - & - & - & - & - & - & - & - & - & - & - \\
Other & 14 & 167 & 5 & 176 & 8 & 53 & 5 & 56 & 8 & 61 & 4 & 65 & 1 & 24 & - & 25 & 1 & 10 & - & 11 & - & 2 & - & 2 & - & 1 & - & 1 \\
Unknown & 2 & 23 & – & 25 & 3 & 22 & 1 & 24 & 21 & 288 & 2 & 307 & – & 26 & – & 26 & 20 & 132 & 3 & 149 & 4 & 9 & – & 13 & – & – & – & – \\
\bottomrule
\end{tabular*}
}
\endgroup
}

\label{tab:category_market_7markets_crypto_misuse}
\end{table*}

\vspace{2mm}
\noindent\textbf{(II) Language-Level Analysis of Misuse.}
 We examine how programming languages influence the likelihood of cryptographic misuse. \autoref{tab:lang_market_7markets_crypto_misuse} summarizes the distribution of programming languages across different MCP markets, along with the presence of cryptographic usage and identified misuses.  To avoid skewed interpretations, when we analyze the data, we exclude extreme cases such as Java and Rust, where the absolute number of servers is too small to draw meaningful conclusions. The results highlight two points: (1) Python’s misuse density is consistently higher across markets, marking it as a critical focus for remediation; and (2) markets with more Python adoption (e.g., Smithery, Mcpmarket) also report higher overall misuse counts, suggesting a compounding effect between language ecosystem characteristics and market curation practices. For example, in Mcpmarket, Python accounts for 89 crypto-enabled servers with 30 misuses, while JavaScript has a larger base (112) but slightly fewer misuses (17). Interestingly, this trend repeats in Smithery Registry. This suggests that although JavaScript remains the most widely used language, Python implementations are disproportionately prone to misuse, likely due to its extensive but unevenly vetted cryptographic library ecosystem.
 \begin{wrapfigure}{r}{0.51\linewidth}
  \centering
  \includegraphics[width=\linewidth]{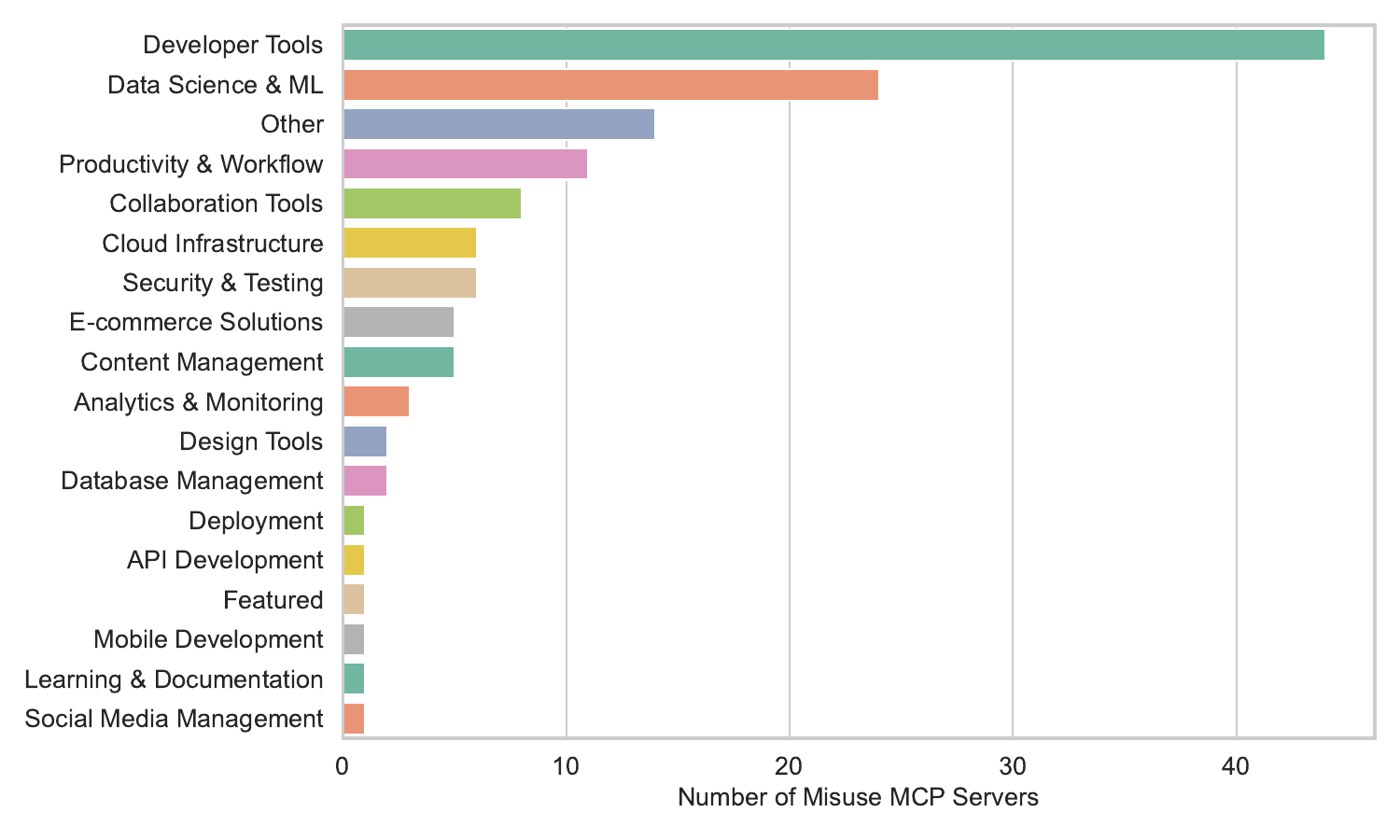}
  \caption{MCP Servers with Crypto Misuse by Category}
  \label{img:MCP_misuse_by_category}
\end{wrapfigure}


\vspace{2mm}
\noindent\textbf{(III) Category-Level Analysis of Misuse.}
We now turn to the analysis of misuse from the perspective of functional categories.
As shown in \autoref{img:MCP_misuse_by_category}, the distribution of misuses is highly uneven, with clear concentration in certain categories. Developer Tools and Data Science \& ML stand out as the two dominant categories, together contributing more than half of the observed misuses. This pattern is not surprising, since these categories often involve direct handling of sensitive data, code execution, or cryptographic operations, which increases both the need for crypto and the likelihood of insecure implementation. In contrast, categories like Learning \& Documentation, Social Media Management, and Mobile Development show only minimal misuses. Tools in these areas are often built upon external APIs or existing frameworks, reducing the chance of developers implementing cryptographic functionality themselves. As shown in \autoref{tab:category_market_7markets_crypto_misuse}, misuse concentration is both category- and market-dependent. Developer Tools and Data Science \& ML consistently emerge as hotspots, while areas like Game Development, Design Tools, and Mobile Development show almost no cases, likely due to limited crypto use or reliance on secure external APIs.





\begin{figure*}[htb]
  \centering

  \begin{subfigure}[t]{0.47\textwidth}
    \centering
    \includegraphics[width=\linewidth]{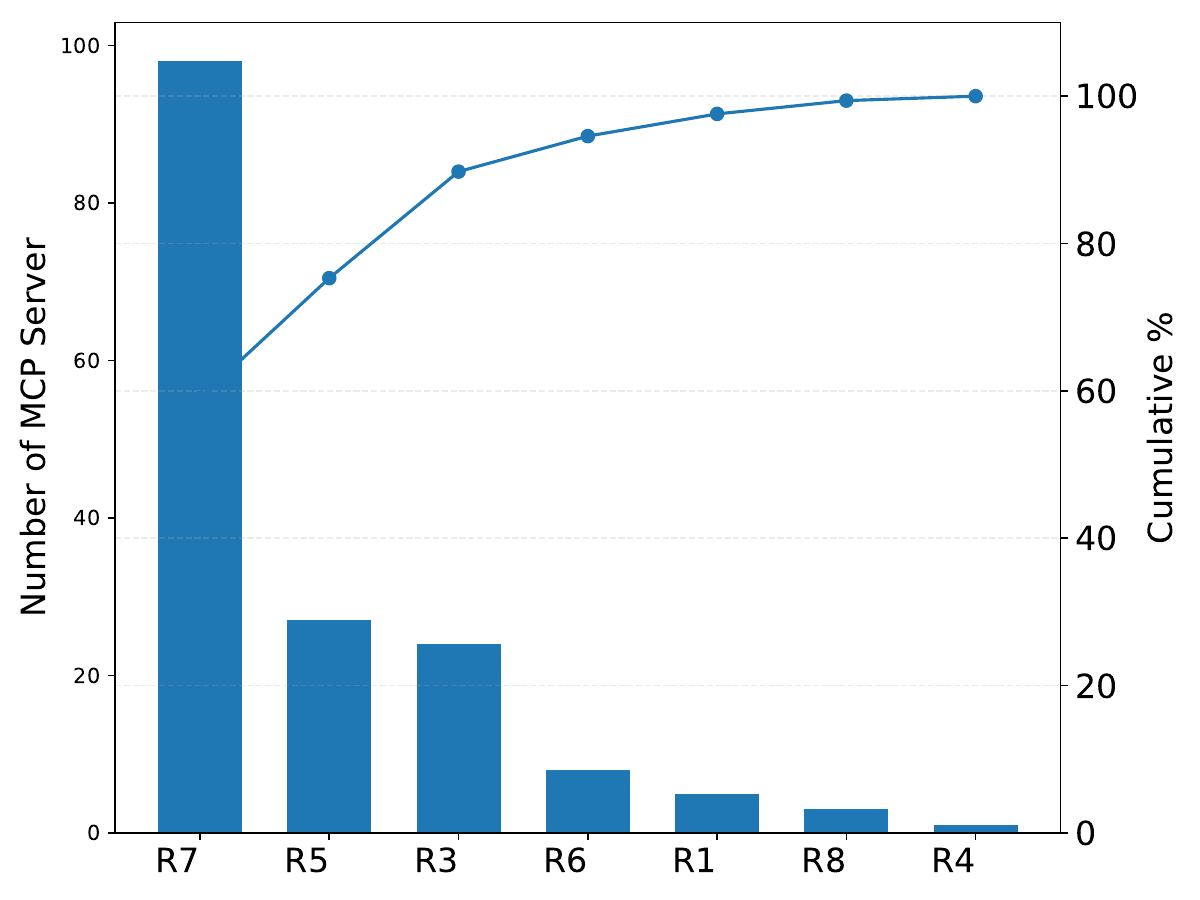}
    \caption{MCP Servers with Crypto Misuse by Rule}
    \label{img:MCP_rule}
  \end{subfigure}
  \hfill
  \begin{subfigure}[t]{0.47\textwidth}
    \centering
    \includegraphics[width=\linewidth]{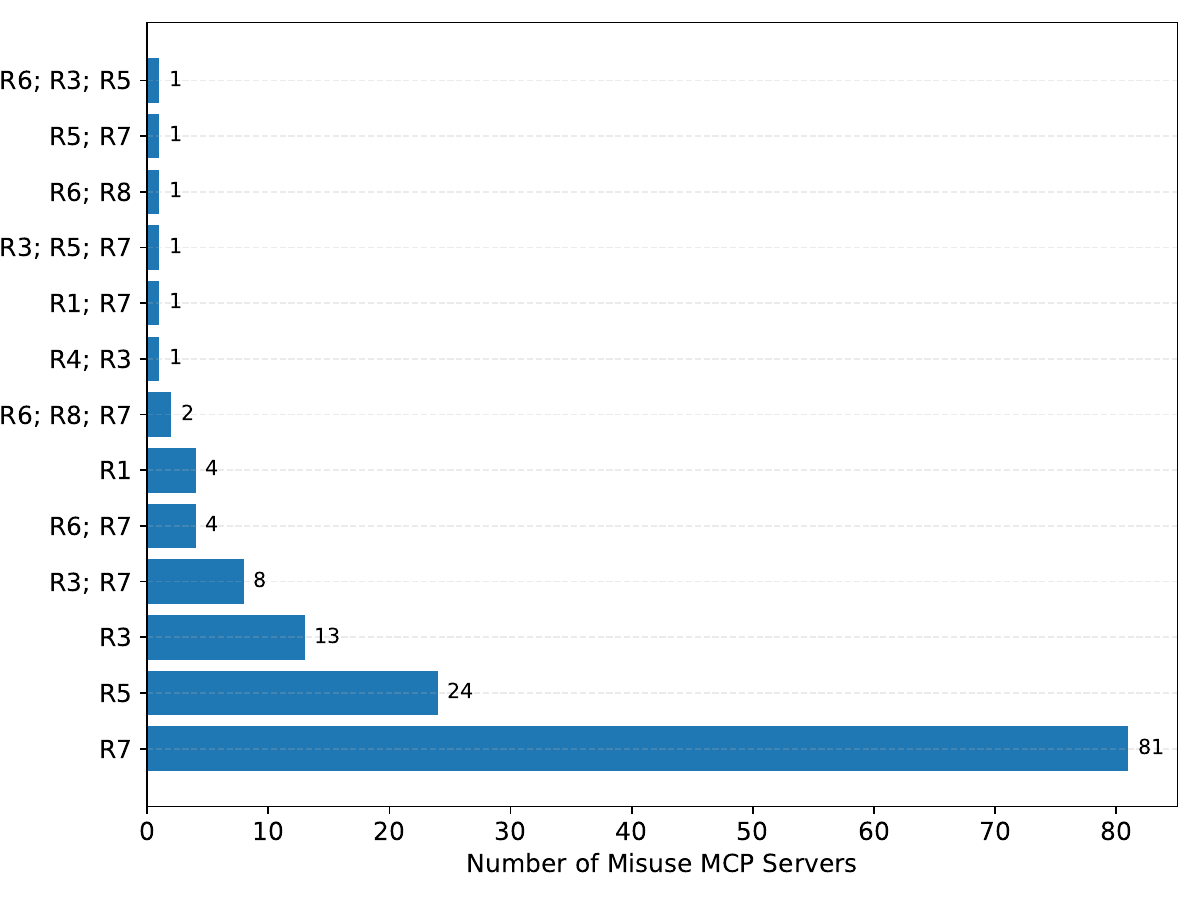}
    \caption{Distribution of Multi-Rule Combinations and Single-Rule Violations in MCP Crypto Misuse}
    \label{img:MCP_rule_combine}
  \end{subfigure}

  \vspace{0.5em}  

  \begin{subfigure}[t]{0.47\textwidth}
    \centering
    \includegraphics[width=\linewidth]{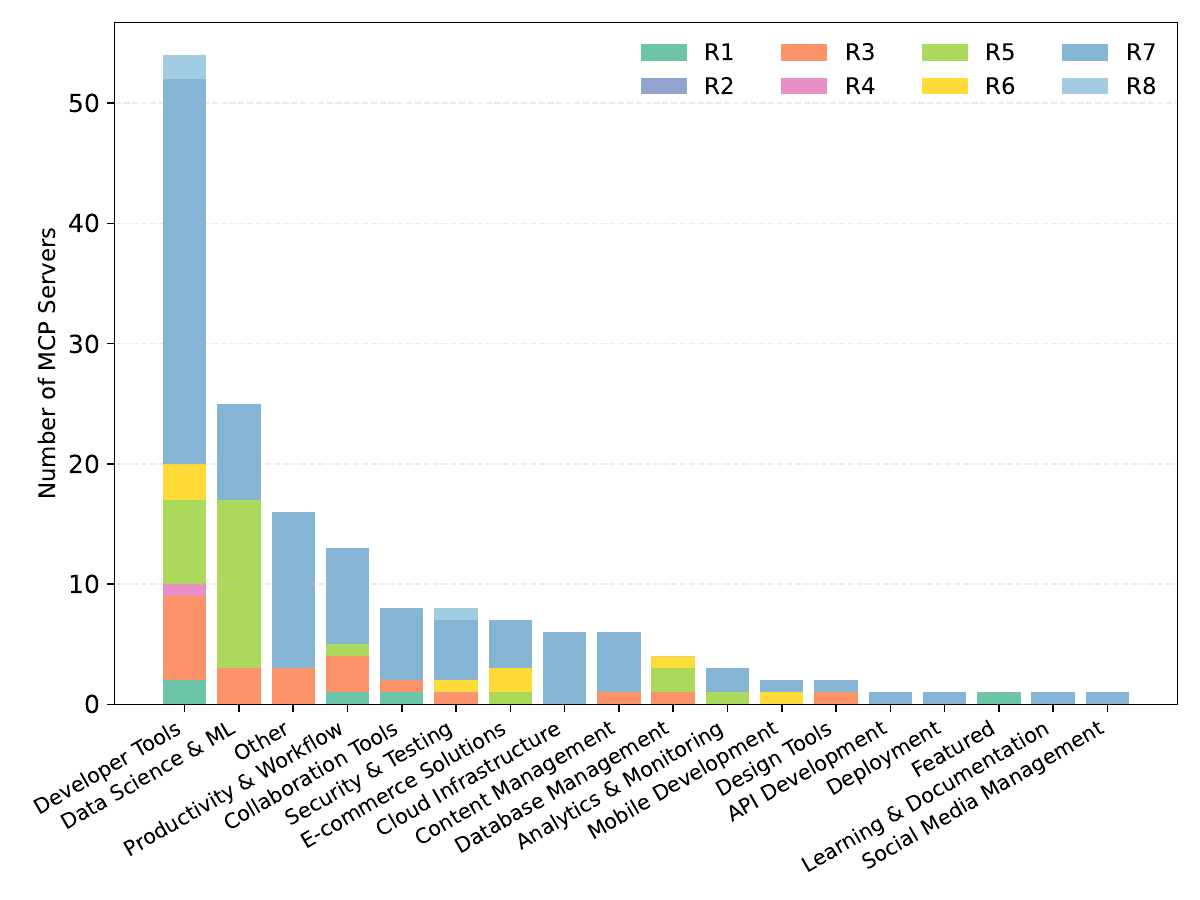}
    \caption{MCP Servers Triggering Rules in Each Category}
    \label{img:MCP_misuse_rule_category}
  \end{subfigure}
  \hfill
  \begin{subfigure}[t]{0.47\textwidth}
    \centering
    \includegraphics[width=\linewidth]{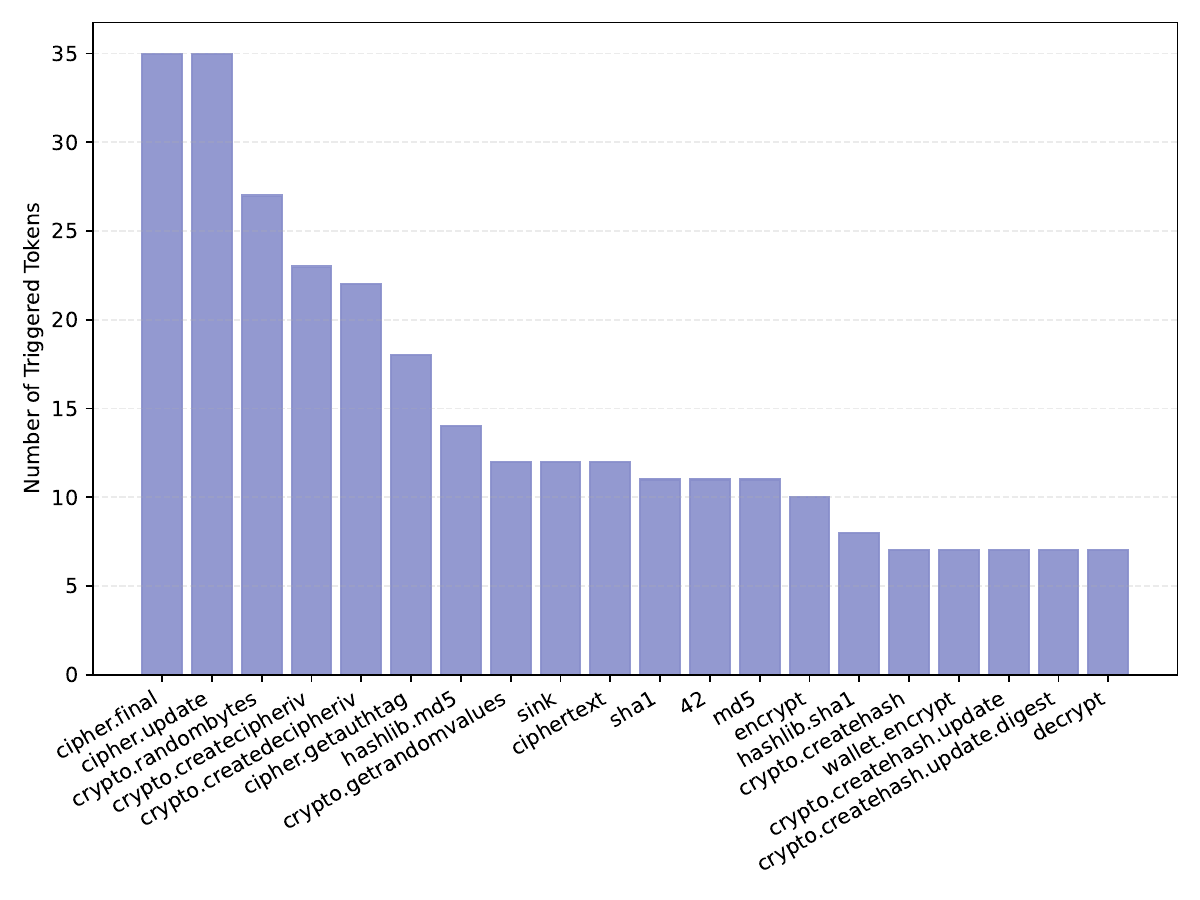}
    \caption{Most Frequent Tokens Triggering Crypto Misuse}
    \label{img:MCP_misuse_token}
  \end{subfigure}

  \caption{Overview of MCP Rule Misuse Patterns}
  \label{fig:mcp_misuse_4plots}
\end{figure*}
\vspace{2mm}
\noindent\textbf{(IV)
Rule-Level Analysis of Misuse.} We examine the distribution of misuses across the eight detection rules (R1–R8). As shown in \autoref{img:MCP_rule}, misuses are concentrated in a few categories, with R7 (Missing Integrity Protection) the most prevalent: servers often apply encryption without authentication (e.g., AES in CBC/CTR mode without MAC/GCM), leaving ciphertext vulnerable to undetected tampering. R3 (Weak Hash Functions) and R5 (Static PRNG Seeds) also stand out, reflecting continued reliance on MD5/SHA-1 and predictable randomness. Importantly, misuses rarely occur in isolation: \autoref{img:MCP_rule_combine} shows that the combination of R3 and R7 (weak digests together with missing integrity checks) and the combination of R6 and R7 (insecure block modes with missing authentication) significantly amplify the attack surface. 
The joint rule–category distribution in \autoref{img:MCP_misuse_rule_category} further indicates clustering within specific domains. Developer Tools exhibit the broadest misuse spectrum (R1–R7), dominated by R7, R5, and R3, while Security \& Testing tools paradoxically also suffer from R7, R5, and R6, often due to static configurations or simplistic client–server protocols. These findings highlight a systemic trade-off of prototyping speed over sound cryptographic design. Finally, token-level analysis (\autoref{img:MCP_misuse_token}) links these rules to concrete coding practices: \texttt{cipher.update/cipher.final} without \texttt{cipher.getAuthTag} drive most R7 cases, while \texttt{hashlib.md5} and \texttt{hashlib.sha1} account for the bulk of weak-hash misuses, confirming that outdated primitives remain pervasive despite long-standing deprecation.

\subsection{Case Study and Security Implications}

\paragraph{Case Study 1: Leaked LLM API Keys.}
We identified MCP servers that embed hard‑coded API keys for LLM services, manifesting a clear violation of R1 (Fixed Key / API Key). For instance, the \textit{Excel Master MCP Server} includes a plaintext Gemini key, and the dify‑for‑dsl MCP Server defines a static \texttt{API\_KEY} (\autoref{lst:api_config}). These exposed keys remain active and usable, allowing unauthorized actors to immediately exploit them. Given that Gemini API usage is billed per token processed (for example, the Gemini 2.5 Pro model charges \$1.25 per million input tokens and \$10 per million output tokens~\cite{gemini_pricing_2025}) such misuse can lead to substantial financial losses, even if the unauthorized usage is modest. If a leaked key is used at scale or inserted into automated scripts, the costs could quickly escalate into hundreds or thousands of dollars.

\begin{figure*}[htbp]
\centering

\begin{minipage}[t][5.2em][t]{0.49\textwidth}  
\begin{minted}[fontsize=\scriptsize, linenos, breaklines, numbersep=3pt]{python}
HOST = '127.0.0.1'
HOST2 = '192.168.2.2'
PORT = 15012
API_KEY = 'AIzaSyD**********************7n0B7nSgCS9U'
PROXY = 'http://127.0.0.1:7897'
\end{minted}
\end{minipage}
\hfill
\begin{minipage}[t][5.2em][t]{0.49\textwidth}  
\begin{minted}[fontsize=\scriptsize, linenos, breaklines, numbersep=3pt]{python}
# Configure Gemini
genai.configure(api_key='AIzaSyAe***********************IBsH1zn4')
model = genai.GenerativeModel('gemini-2.0-flash-001')  # Use flash model for structured extraction
\end{minted}
\end{minipage}
\caption{MCP Server Configuration (Hardcoded API Key): Excel Master MCP Server (left) vs. dify‑for‑dsl MCP Server (right)}
\label{lst:api_config}
\end{figure*}

\vspace{1mm}
\paragraph{Case Study 2: Insecure Crypto MCP Server.}  
The \textit{Crypto MCP Server} exposes cryptographic primitives as standardized MCP tools, making them callable by IDEs or AI assistants. However, one of its tools implements {DES in ECB mode} (\autoref{lst:crypto}), a configuration widely considered insecure. DES’s small key size makes brute-force feasible, while ECB mode leaks plaintext structure by encrypting identical blocks deterministically. Exposing such a tool through MCP magnifies the danger into a \textit{supply-chain vulnerability}, where insecure primitives can be reused by many downstream applications, severely undermining confidentiality and integrity guarantees.  

\begin{figure}[htbp]
\begin{minted}[fontsize=\scriptsize,breaklines, numbersep=3pt]{typescript}
const encrypted = CryptoJS.DES.encrypt(message, keyHex, {
  mode: CryptoJS.mode.ECB,
  padding: CryptoJS.pad.Pkcs7,
});
\end{minted}

\caption{DES with ECB Mode in Crypto MCP Server}
\label{lst:crypto}
\end{figure}

\vspace{1mm}
\paragraph{Case Study 3: Weak Hash in 1Panel MCP Server.}  
The \textit{1Panel MCP Server} provides automated deployment capabilities but implements a flawed authentication scheme. Each request header includes a token computed as \texttt{MD5("1panel" + apiKey + timestamp)} (\autoref{lst:getAuthHeaders}). The reliance on \textbf{MD5}, a weak hash function, enables efficient offline brute-forcing if a token or API key is leaked. Because MD5 is susceptible to preimage and collision attacks, this design allows adversaries to forge valid tokens and compromise the deployment pipeline. The impact is severe: an attacker could bypass authentication and gain control over the deployment process, directly threatening system integrity.  
\begin{figure}[htbp]
\begin{minted}[fontsize=\scriptsize, breaklines, numbersep=3pt]{typescript}
getAuthHeaders() {
  const timestamp = Math.floor(Date.now() / 1000).toString();
  const content = `1panel${this.apiKey}${timestamp}`;
  const token = crypto.createHash("md5").update(content).digest("hex"); 
  return {
    "1Panel-Token": token,
    "1Panel-Timestamp": timestamp,
    "Accept-Language": this.languageCode,
  };
}
\end{minted}

\caption{Token Generation Function \texttt{getAuthHeaders()} in 1Panel MCP Server}
\label{lst:getAuthHeaders}
\end{figure}

	\section{Discussion}
\label{section:discussion}
 
\noindent\textbf{Limitations.}
Although \sysname demonstrates strong capabilities in systematically uncovering cryptographic misuses across heterogeneous MCP servers, several limitations remain. First, our analysis pipeline is primarily static and relies on intermediate IR construction together with taint propagation. As a result, dynamic aspects of MCP servers (such as IVs generated at runtime from environment randomness or keys derived through user interactions) may not be fully captured, occasionally leading to false negatives. Second, our rule set, while covering eight representative misuse categories, is still bounded by pre-defined patterns. Although such semantic gaps can obscure misuses (e.g., crypto wrapped in custom utility functions), in practice these cases are relatively rare compared to the broader misuse patterns we observed.Third, \sysname focuses on code-level misuse and does not yet cover deployment issues (e.g., insecure transport, weak dependencies, outdated libraries), though these can also undermine MCP security but are extendable in future work.

\vspace{1.5mm}
\noindent\textbf{Root Causes of Misuse in the MCP Ecosystem.}
A central question raised by our study is why cryptographic misuses (19.7\%) are so pervasive in the MCP ecosystem. 
MCP provides only minimal safeguards and lacks authenticity or confidentiality guarantees, pushing developers to implement custom crypto prone to classic pitfalls.
Second, the heterogeneity of MCP implementations intensifies this challenge: servers span more than ten programming languages, each with distinct API conventions, implicit defaults, and pitfalls, which complicates the enforcement of uniform secure practices.  
Third, the plugin-style architecture of MCP tools fosters weak coupling and implicit execution paths. Because LLMs dynamically orchestrate functions at runtime, secure and insecure utilities may coexist, and insecure compositions can emerge only under certain prompt-driven workflows. Finally, the developer-centric and market-driven nature of MCP encourages rapid prototyping and mass publication of servers, where functionality and interoperability often take precedence over cryptographic rigor. 

\vspace{1.5mm}
\noindent\textbf{Ethical Considerations.} This work focuses on analyzing publicly available MCP server implementations to detect cryptographic misuse. We carefully considered the ethical implications of our methodology and findings, guided by the principles outlined in the Menlo Report and subsequent discussions on ethical frameworks in computer security research.

\begin{itemize}
\item  \textbf{Impacts and Ethical Principles}. Following the principle of Beneficence, we designed our methodology to maximize positive outcomes (identifying and mitigating systemic weaknesses) while minimizing potential harms. Respect for Persons was upheld by avoiding any collection of private user data; our analysis was limited to open-source code and public registries. Justice guided our focus on a broad and representative set of MCP servers, ensuring that security improvements benefit the entire ecosystem rather than a narrow subset. Finally, Respect for Law and Public Interest was maintained by operating only on publicly available artifacts and adhering to repository terms of service.

\item \textbf{Harms and Mitigations}. We acknowledge potential harms such as reputational damage to developers whose insecure implementations are identified, or the possibility that adversaries could misuse our findings to locate exploitable systems. To mitigate these risks, we (i) did not disclose vulnerabilities in a way that enables direct exploitation, (ii) reported critical cases to responsible parties.  For example, we responsibly reported exposed keys to affected vendors. In several cases, vendors promptly revoked or deprecated the leaked credentials following our disclosure. To confirm these actions, we relied on safe, non-inference validation methods such as querying metadata or account-status endpoints~\cite{stackoverflow2023_check_api_key_python,github2024_gpt_researcher_discussion,openai2024_api_key_format}, which allow us to verify whether a key is active without incurring costs or triggering model inference. This approach ensured that our verification process did not itself introduce risk, while providing evidence that remediation steps were indeed taken. 
\end{itemize}

We conclude that the ethical benefits outweigh potential harms: systematically documenting the prevalence and causes of cryptographic misuse in MCP serves the community by highlighting systemic risks and motivating stronger standards, tools, and practices. The decision to publish is supported both by beneficence (maximizing ecosystem security) and by respect for law and public interest (avoiding unlawful or privacy-invasive actions)

	\section{Related Work}

{\bf Cryptographic Misuse.} Recent studies show that cryptographic misuse remains a widespread and persistent threat to software security, even in modern development environments \cite{ami2022crypto,fischer2024challenges,karimova2025model,wang2024cryptody}. There are various types of misuse, such as the use of weak algorithms (e.g., ECB mode \cite{chen2024towards}, MD5/SHA-1 \cite{torres2023runtime,wickert2022fix}, DES/RC4 \cite{li2022cryptogo}) flawed configurations (e.g., fixed IVs \cite{piccolboni2021crylogger}, fixed seeds \cite{xia2025beyond}, or salts \cite{mandal2024belt,zhang2024gopher,gilsenan2023security,kim2024privacy}, and low PBKDF2 iterations \cite{sun2023cryptoeval}). Cryptographic misuse can arise in a variety of domains. For example, Wang et al.~\cite{wang2024cryptody} found that 95\% of 1,431 IoT firmware samples contained misuse. Yu et al.~\cite{ShiYZ0Y0025} observed that mini-applications are also prone to cryptographic misuse~\cite{ChenCWWJSXP25,Zhang0L23}.
Many misuses stem from developers’ misunderstanding of cryptographic APIs. Therefore, researchers proposed static analysis and rule-based detection as effective solutions. Several studies formalized common misuse patterns to facilitate tool development \cite{chen2024towards}, while others highlighted the need to enhance detection capabilities in the context of LLM-generated code \cite{xia2025beyond,fang2024large,he2023large}. Our work differs by providing the first systematic study of cryptographic misuse in MCP.

\vspace{1mm}
\noindent{\bf MCP Security.} MCP is a standardized interface that connects LLMs with external tools by structuring inputs and coordinating multi-source information \cite{karimova2025model,11069802,11105280}. However, recent studies have shown that if an insecure MCP is connected, it can become an exploit path for attackers to control LLM behavior, inject malicious instructions, steal assets, and achieve remote code execution (RCE) \cite{liu2024demystifying,dong2025philosopher}. Specifically, untrusted MCP data sources or tool responses can lead to serious risks such as prompt injection \cite{liu2025datasentinel,liu2024formalizing,shen2024anything}, cross-service prompt stealing \cite{yangprsa}, demystifying RCE manipulation of executors \cite{liu2024demystifying}, and trojanizing plugins \cite{dong2025philosopher}. Furthermore, the complexity of responsibility attribution introduced by long contexts requires enhanced traceability in MCP design \cite{wangtracllm}. Unlike these studies focusing on prompt injection or malicious plugins, we examine misuse of cryptographic operations inside MCP servers, exposing a complementary layer of risk.

	\section{Conclusion}
We introduced \sysname, the first framework for detecting cryptographic misuses in MCP implementations. 
Our large-scale study of 9,403 servers shows that 19.7\% of crypto-enabled MCP servers contain misuses, with risks clustering in specific markets, languages, and categories. Common pitfalls create tangible threats from financial abuse to supply-chain vulnerabilities. These findings demonstrate that misuse in MCP is systemic, underscoring the need not only for better detection but also for automated remediation, developer guidance, and protocol-level defenses to ensure MCP can serve as a secure foundation for the agentic AI ecosystem.


	
	\bibliographystyle{plain}
	\bibliography{reference}
	
\end{document}